\documentclass[useAMS,usenatbib]{mn2e}
\usepackage{amsmath}
\pdfoutput=1
\usepackage[pdftex]{graphicx}
\usepackage{epstopdf}
\usepackage[english]{babel}
\usepackage{subfigure}
\usepackage{array,multirow}
\usepackage{underscore}
\usepackage{natbib}

\def\mnras{MNRAS}
\def\apj{ApJ}
\def\aj{AJ}

\def\aaps{A\&AS}

\def\apjs{ApJS}
\def\araa{ARA\&A}

\title[M87's globular cluster populations]{Galaxy structure from multiple tracers: I. A census of M87's globular cluster populations}
\author[L. J. Oldham and M. W. Auger]{L. J. Oldham\thanks{E-mail: ljo31@ast.cam.ac.uk}, M. W. Auger
\\
Institute of Astronomy, University of Cambridge, Madingley Road, Cambridge CB3 0HA, UK}

\pubyear{2015}

\begin{document}
\maketitle
\setcounter{page}{1}

\begin{abstract}
We present a new photometric catalogue of the rich globular cluster (GC) system around M87, the brightest cluster galaxy in Virgo. Using archival Next Generation Virgo cluster Survey (NGVS) images in the \textit{ugriz} bands, observed with CFHT/MegaPrime, we perform a careful subtraction of the galaxy's halo light in order to detect objects at small galactocentric radii as well as in the wider field, and find 17620 GC candidates over a radius range from 1.3 kpc to 445 kpc with $g <$ 24 magnitudes. By inferring their colour, radial and magnitude distributions in a Bayesian way, we find that they are well described as a mixture of two GC populations and two distinct contaminant populations, but confirm earlier findings of radius-dependent colour gradients in both GC populations. This is consistent with a picture in which the more enriched GCs reside deeper in the galaxy's potential well, indicating a role for dissipative collapse in the formation of both the red and the blue GCs. 
\end{abstract}


\section{Introduction}
The globular cluster (GC) populations of a galaxy hold a wealth of information about the galaxy itself, both past and present. While the frequently observed bimodality in their colours \citep{Zepf1993, Ostrov1993,Gebhardt1999} hints at a non-trivial formation history, with at least two major formation periods and mechanisms at work, their extended spatial distributions make them good probes of the dark matter profile in the halo, which can otherwise prove elusive. In this way, GC dynamics can offer powerful insights into a galaxy's extended mass structure in a way that the much more centrally-concentrated starlight alone cannot.

However, any inference based on a subsample of a galaxy's GCs depends on a proper characterisation of the spatial profiles of the underlying population. For instance, the subsample of GCs in extragalactic systems for which we can obtain reliable spectroscopy tends to be subject to some non-trivial selection criteria, and this can lead to apparent spatial distributions which deviate dramatically from those of the parent populations. Dynamical models of the galaxy based on these spectroscopically-determined distributions rather than the true underlying ones can therefore result in very different conclusions. In the same way, as GCs generally populate their host galaxies with surface density profiles that fall off rapidly with galactocentric distance -- often characterised, for instance, by S\'ersic profiles -- it is important that this characterisation of the underlying population is informed by GCs as close-in to the galaxy's centre as possible; this way, the innermost slope of the profile can be constrained with much less uncertainty. These GC populations (and planetary nebula populations, if present in significant numbers), with their distinct spatial and kinematic signatures, can then be used in dynamical models as independent tracers of the gravitational potential, providing larger-radius constraints on the galactic structure which are complementary to those from the starlight.

The massive elliptical M87, located at the centre of the Virgo cluster, is an ideal subject for dynamical GC projects such as these. This brightest cluster galaxy (BCG) hosts the largest collection of GCs in the local Universe, with estimates as large as N$\sim$12,000 \citep{McLaughlin1994, Tamura2006a, Durrell2014}. Its large GC population was first recognised by \citet{Baum1955}, in a study which compared the GC luminosities with those of M31 for use as a distance indicator. Since then, its GCs have been the target of a number of observational programmes, and recent work has conclusively shown that there are at least two distinct populations of GCs that are separated both in colour space and in their radial profiles, with the blue GCs extending significantly further than the red GCs \citep[e.g.][]{Harris2009}. These multiple populations are important dynamical tracers of the potential well at different radii, and several groups have recently exploited the large catalogue of spectroscopic GC velocities from \citet{Strader2011} to place constraints on the dark matter profile and stellar mass-to-light ratio of M87 \citep{Agnello2014,Zhu2014}.

However, this high-quality spectroscopy has yet to be coupled with a radially complete, publicly available photometric catalogue of the kind needed to properly characterise the underlying populations. Indeed, while a number of photometric catalogues exist, most have only partial radial coverage, and care must be taken to relate these catalogues when the observations are made using different filters and instruments. The HST catalogue of \cite{Peng2009}, for instance, is extremely deep and complete out to a galactocentric radius of $\sim$ 6.5 kpc in the ACS F606W and F814W filters, while NGVS provides archival catalogues covering a large extent of the Virgo cluster in the CFHT/MegaPrime $ugriz$ bands, but with the regions immediately surrounding all bright objects (including M87) removed. The impetus for this paper, then, is to bridge the gap and compile a comprehensive, extensive catalogue in a single filter system, by going back to the original NGVS images to carefully model and subtract the galaxy light and produce catalogues with more complete radial coverage. These can then be used to robustly characterise the GC distributions, acting as a springboard for future work. 

The paper is structured as follows: in Section 2, we introduce the dataset and explain our background subtraction. In Section 3, we present our photometry; Section 4 describes our GC selection methods and Section 5 our inferred distributions. We discuss our findings in Section 6 and summarise in Section 7. Throughout the paper, we adopt a distance to the centre of M87 of $D_L$ = 16.5 Mpc, corresponding to an angular scale of 83pc/$''$. All magnitudes are in the AB system.

\section{Data \& reduction}
We downloaded stacked NGVS images in the region around M87 from the Canadian Astronomy Data Centre (CADC). NGVS imaged a total of 104 deg$^2$ within the virial radii of Virgo's A and B sub-clusters in the $ugriz$ bands of the MegaPrime instrument on the Canada-France Hawaii Telescope (CFHT), over 200 nights between 2009-2014 \citep[full details can be found in ][]{Ferrarese2012}. The images available for download have a pixel scale of 0.187$''$/pixel and have been pre-processed using the (NGVS-tailored) Elixir-LSB pipeline -- which includes bad pixel masking, bias and overscan subtraction, flat fielding and the removal of scattered light -- and then photometrically and astrometrically calibrated and stacked. They are accompanied by exposure time and bad pixel maps and preliminary source catalogues obtained using SExtractor. We downloaded images for the four fields covering M87 and its surroundings: in the NGVS file-naming system (in which the first number refers to the RA offset from the survey centre and the second to the Dec equivalent), these are the +0+0,+0+1,-1+0 and -1+1 tilings. As noted in the Introduction, the part of the image immediately surrounding M87 is extremely bright due to M87's stellar profile, making source detection and measurement incomplete and unreliable out to a radius of 6 kpc, and this region has also been removed from the NGVS catalogues. 

\begin{figure*}
  \centering
  \subfigure{\includegraphics[trim=120 0 140 0,clip,scale=0.51]{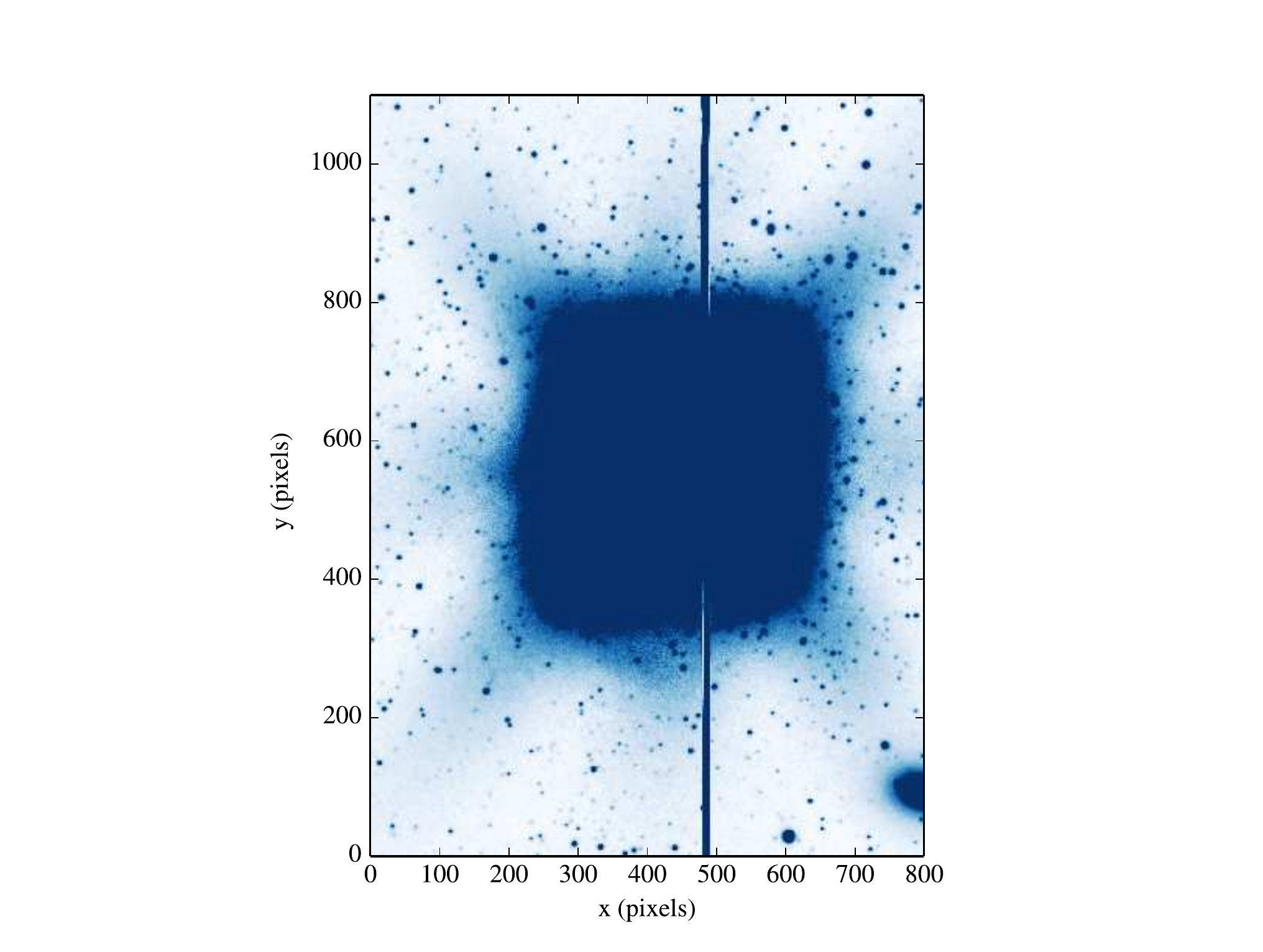}}\hfill
  \subfigure{\includegraphics[trim=120 0 140 0,clip,scale=0.51]{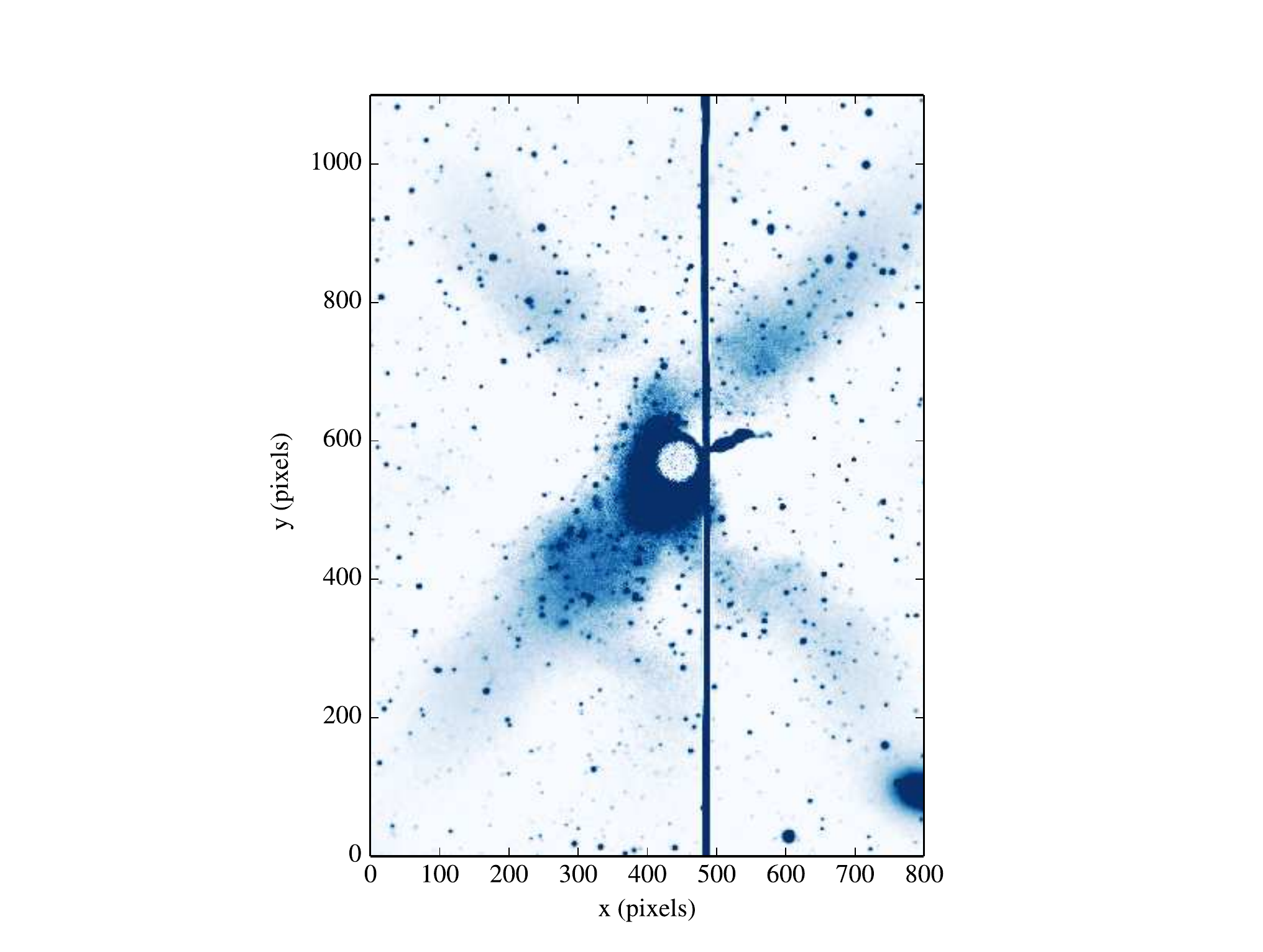}}\hfill
  \subfigure{\includegraphics[trim=120 0 140 0,clip,scale=0.51]{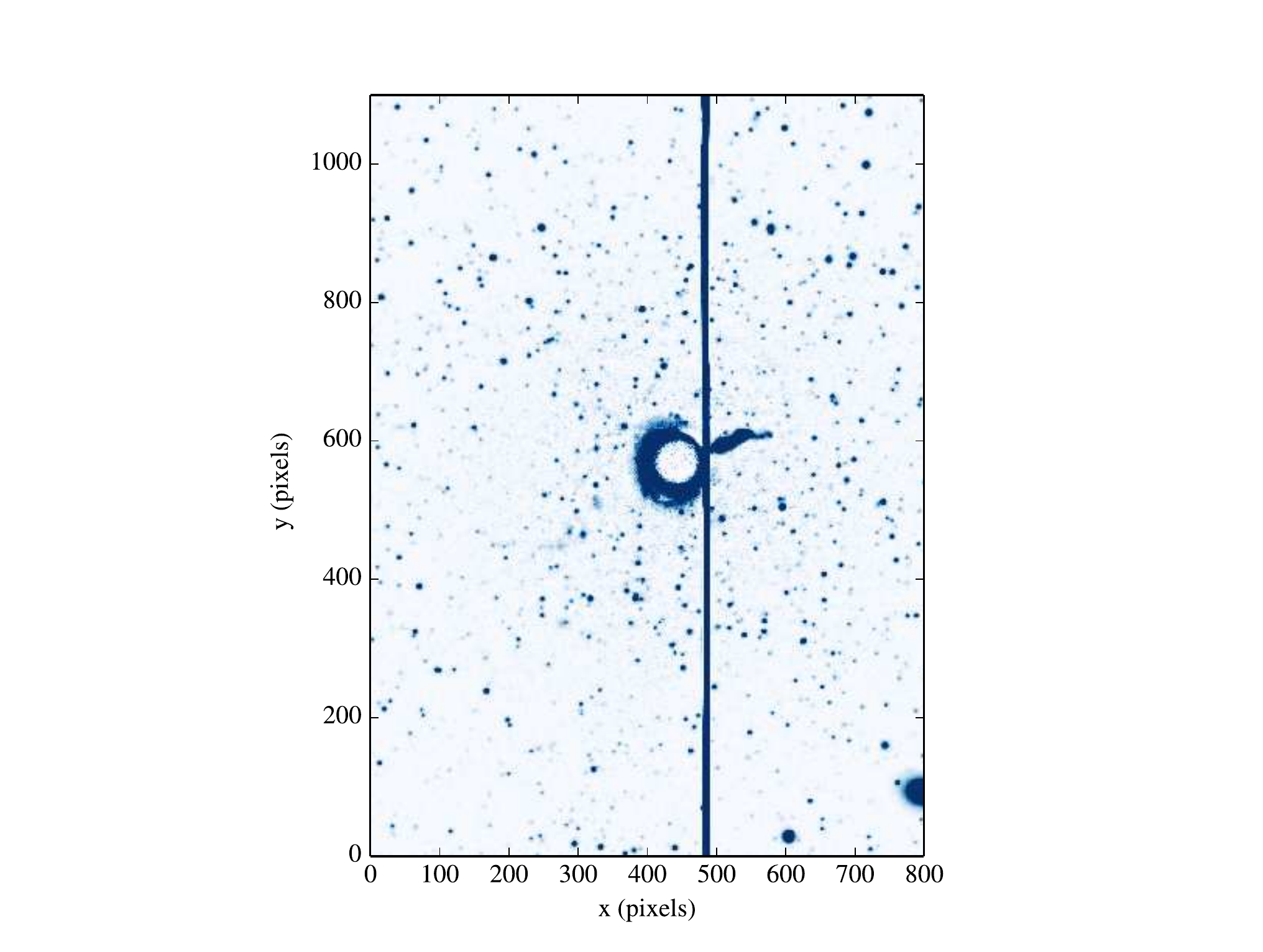}}\hfill
\caption{Step-by-step light subtraction in the $r$-band. Left: In the original NGVS image, the GCs close to M87 are totally dominated by the stellar light, making photometry unfeasible. Centre: Subtracting a spline model of the light in the radial direction helped to reveal the central region, but the lack of azimuthal symmetry in the original image resulted in a residual X-shaped pattern. Right: After the subtraction of a smooth Fourier-filtered background, the inner region of M87 is much cleaner. The circle in the very centre represents the saturated pixels, and is surrounded by a small bright ring which we were unable to totally eliminate, but whose size we substantially reduced; the saturated columns to the right of M87 are bleeding down from a star at higher declination. All three images are shown on the same scale.}
\label{fig:bgsub}
\end{figure*}

Careful modelling and subtraction of the stellar light distribution allowed us to significantly improve the situation. We first performed an object detection step to avoid removing light from globular cluster candidates and thereby underestimating their flux. This was accomplished by first subtracting a heavily median-filtered version of the image from the original, then applying a clipped mean/variance filter to flag all pixels with flux 2$\sigma$ above the local noise level, and finally applying a series of erosion and dilation filters to the flag image to remove noise and expand the mask around real objects. This mask was then applied to the original image and we fitted a smooth spline to the light profile in the radial direction, which we subtracted. This removed a significant component of M87's stellar light, as can be seen by comparing the first two panels of Figure~\ref{fig:bgsub}, but as the distribution of the light is not purely radial, this left the image with the X-shaped pattern that is prominent in the central panel. To remove this effectively, we Fourier-filtered the high frequencies from the intermediate image to construct a smooth background, which could again be subtracted. The difficulty here is that Fourier filtering requires a real-space image with no gaps, i.e. no masked sources, so we iterated this procedure a number of times, each time updating the source mask and using the current best smooth background image to fill in the masked pixels and so converge to an acceptable approximation of the background.

As can be seen in the right-hand panel of Figure~\ref{fig:bgsub}, the result of this step was a much cleaner image down to $\sim$1.1 kpc, with a clear jet shooting off to the right and a large number of GCs revealed close to the galaxy's centre. In all except the $u$-band (where the galaxy is not very bright), the very central region proved impossible to model due to saturation, and, having made this region as compact as possible, we masked it in the subsequent analysis.

\section{Photometry}
\label{sec:phot}
We used SExtractor and PSFex \citep{Bertin1996} to detect and measure sources. For these routines to work in conjunction, SExtractor must be run first to provide a set of small images (`vignettes') for each detection. These vignettes are then used by PSFex to construct a model of the point-spread function (PSF) which can vary across the field of view, and which is fed into a final run of SExtractor to measure PSF-fitted magnitudes. This three-step process is preferable to running SExtractor alone when the field is crowded, as it enables a greater degree of deblending than could otherwise be achieved. SExtractor can also use a weight map alongside the detection and measurement images to deal with variations in noise, and here we found that the use of the NGVS exposure time maps was not sufficient to handle the increased shot noise in the innermost regions of the galaxy as compared to the wider field. To account for this properly, we combined the exposure time maps with a second set of maps quantifying the relative pixel-to-pixel variance. We then ran the SExtractor-PSFex-SExtractor routine for each of the five bands and each of the four fields, using the $g$-band as the detection image as this had the most independent detections and so allowed us to obtain as complete a catalogue as possible. We added the uncertainties on the magnitudes output by SExtractor in quadrature with a systematic uncertainty of $0.02$ mags to account for uncertainty in the photometric zeropoints. The use of a detection image made the merging of the different bands trivial; we also merged the catalogues from the different fields, removing duplicates in the overlapping regions by discounting any objects centred within 1.5 pixels of one another.

To remove regions of the image with unreliable photometry -- primarily, those close to the bleed trails of very bright stars and their pupil ghosts -- we made use of some of the other parameters output by SExtractor, identifying all objects fulfilling the following criteria in the $g$-band, and applying a dilation filter to a map of their positions to create a mask:
\begin{enumerate}
 \item \textit{FLUX_RADIUS \textgreater 10 pixels}: SExtractor measures a half-light radius based on the flux inside a (user-defined) circular aperture, which can be used to identify objects likely to be extremely bright foreground stars
  \item \textit{MAG_AUTO \textless 16 mags}: this magnitude measurement, made using flexible apertures centred on each detection prior to PSF-fitting, is preferable to the PSF-fitted magnitude for identifying bright extended objects, whose fluxes are likely to be underpredicted by the PSF model.
\end{enumerate}

Based on this masking, we calculated the fractional effective unmasked area $\frac{A_{eff}}{A}(R)$ of our field of view as a function of galactocentric radius and fitted it with a spline; this is important for relating measured number counts to physical surface densities. Explicitly, the observed and intrinsic number counts are simply related by
\begin{equation}
\begin{split}
n_{int}(R) &= n_{obs}(R) / \frac{A_{eff}}{A}(R) \\
N_{int}(R) &= N_{obs}(R) / \frac{A_{eff}}{A}(R) 
\end{split}
\end{equation}
for surface number density $n$ and total number $N$.

To test our photometry, we cross-matched our objects with those in the NGVS and HST catalogues. When comparing with NGVS, a small amount of scatter is to be expected due to differences in our detection methods -- the NGVS catalogues were generated using the exposure-time weight maps only, and without modelling the PSF -- but the overall scatter is just $\sim$0.017 mags down to 24th magnitude. We performed a similar comparison  with the HST/ACS B,V,I photometry in the catalogue of \cite{Peng2009}, this time selecting GC candidates as explained in Section~\ref{sec:gcc} and calculating synthetic photometry based on a mixture model of four single stellar populations (SSPs), at different ages and metallicities to reflect the bimodal nature of the GC population. The scatter in this case is still consistent with experimental uncertainty at $\sim$ 0.03 mags, especially given the simplicity of our four-component SED model. We also used the HST image to measure the completeness of our catalogue, given its depth and extremely high resolution. Using our best-fit synthetic photometry, we confirmed that our catalogue remains complete down to 24$^{\text{th}}$ magnitude, with no dependence on radius outside the masked central $\sim1$~kpc region.

As a further check that the background subtraction around M87 had not biassed our photometry, we ran a series of simulations, synthesising stellar objects with Gaussian profiles and known magnitudes and inserting them into the original pre-subtraction image, then running the subtraction procedure and comparing the magnitudes output from SExtractor with their known magnitudes. We did this for a total of 100 sources, implanting groups of 10 to avoid dramatically changing the density, and found that SExtractor managed to consistently reproduce the magnitudes to within a maximum difference of 0.04 mags. We also ran a similar experiment to test SExtractor's detection efficiency in the central regions around M87, inserting synthetic objects exclusively in this area; SExtractor was able to detect and recover unbiassed photometry for all sources. 
 
\section{Globular Cluster Candidates}
\label{sec:gcc}
\begin{figure}
  \centering
  \includegraphics[width=0.5\textwidth]{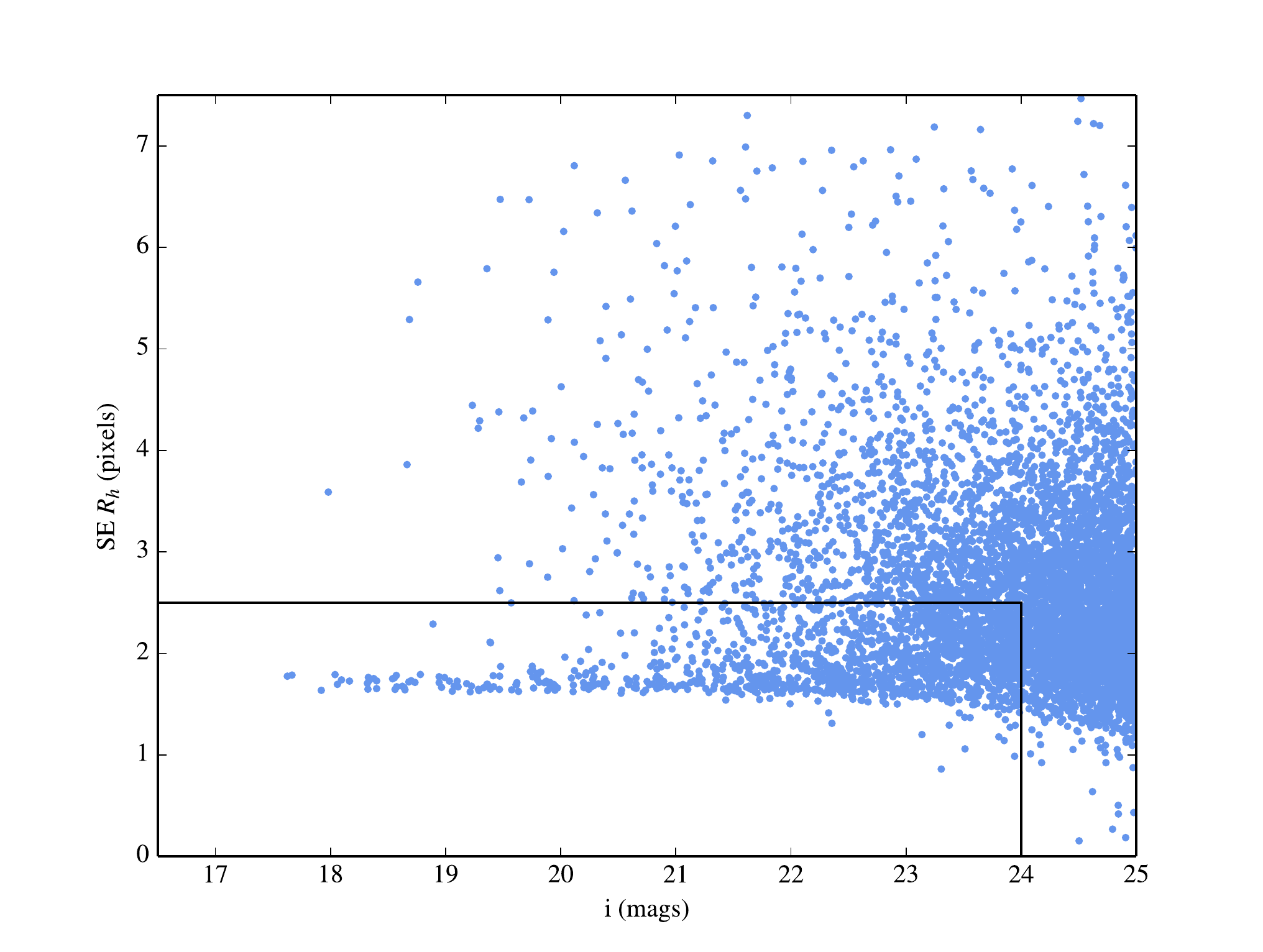}
  \caption{Point source selection using SExtractor's effective radius parameter: SExtractor measures a half-light radius for each detection based on its fixed-aperture magnitudes, which, in the case of point sources, can be used as a proxy for the PSF. Point sources therefore lie on a distinct branch of small radius with low scatter. For each field, all objects outside the stellar selection box (defined using the $g$-band image) were rejected from the sample. At the faint end of the box, we chose our magnitude cut-off such that our catalogue depth would be comparable to the turnover magnitude of the GC populations and would include the main population of \protect\cite{Strader2011}'s kinematic sample. Clearly, this choice of magnitude cut-off is a compromise between the detection of fainter GCs and the amount of contamination in the catalogue.}
\label{fig:GCsel1}
\end{figure}

\begin{figure*}
  \centering
  \subfigure{\includegraphics[trim=20 0 20 0,clip,scale=0.46]{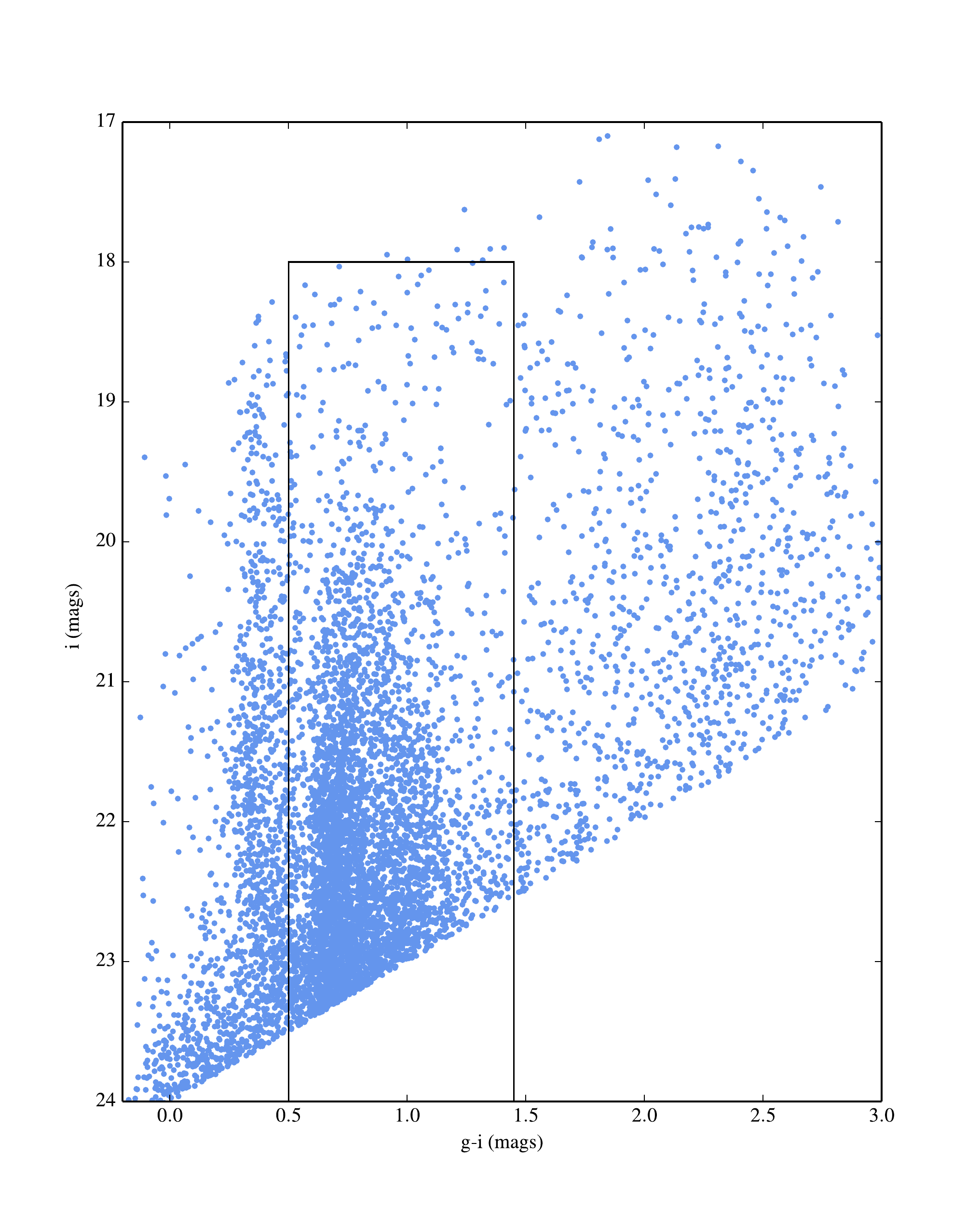}}\hfill
  \subfigure{\includegraphics[trim=20 0 20 0,clip,scale=0.46]{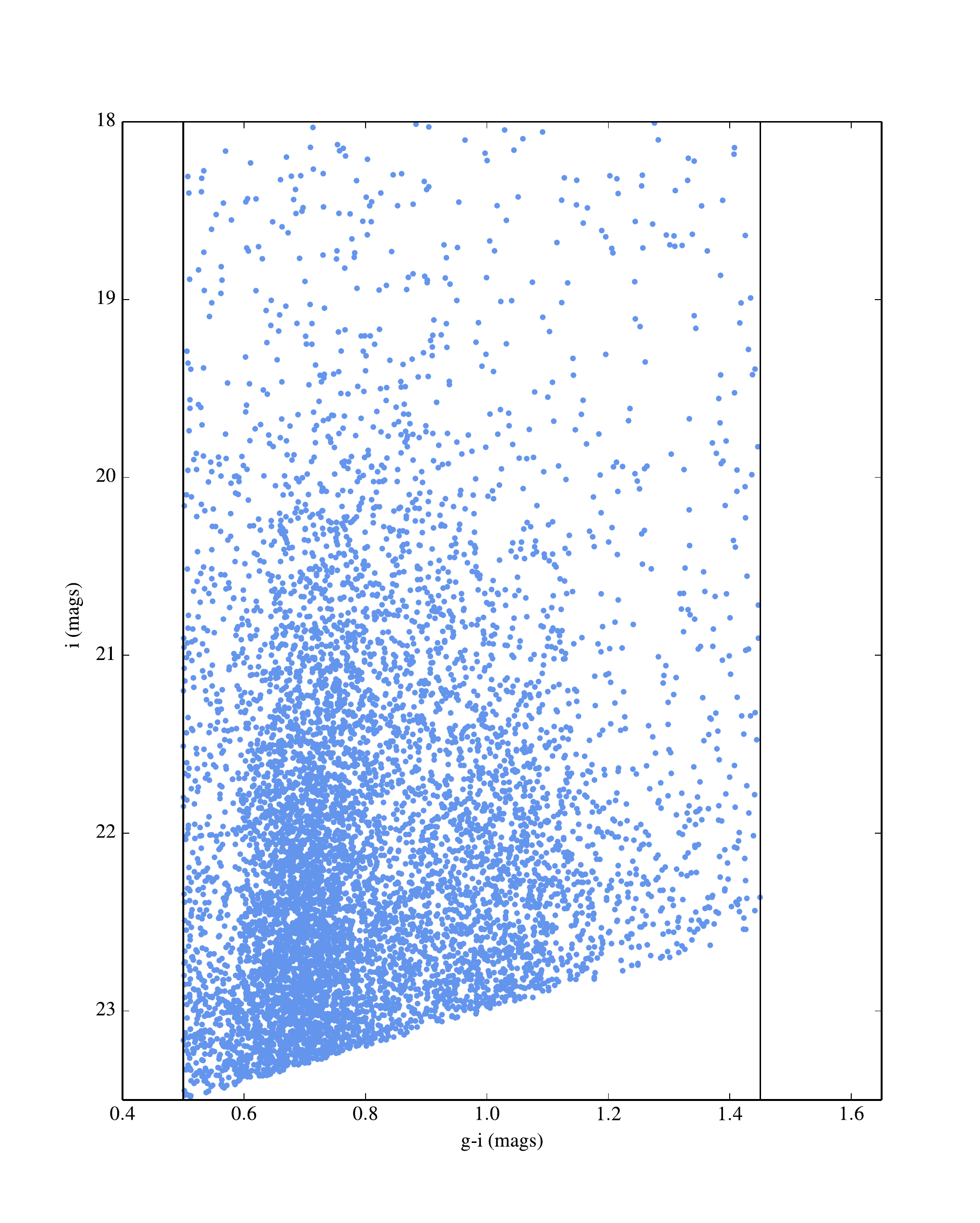}}\hfill
\caption{Colour selection: M87's GCs are known to be bimodal in colour space, as can be seen particularly clearly in the $g-i$ CMD. Left: The distribution of all point sources, with our selection limits overplotted. The bluest population at $g-i\sim$0.4 is made up of MSTO stars in the Milky Way halo, while the extremely red component is due to low-luminosity disk stars. Bracketed by these is the double-peaked GC distribution, although note that stars from the Sagittarius stream also have very similar colours to the blue GC population. Right: The $g-i$ CMD after all colour cuts have been applied. }
\label{fig:GCsel2}
\end{figure*}

The line of sight to M87 is heavily contaminated by stars in the Milky Way halo (and, to a lesser extent, the thick disk) and the Sagittarius stream, and is seen against a backdrop of interloping galaxies. We identified GC candidates according to the sizes, colours and magnitudes of sources via the following steps:

\begin{enumerate}
\item \textit{Selecting point sources}: Plotting magnitude against SExtractor-measured sizes, a clear horizontal distribution of unresolved objects emerges. Treating each field separately to allow for variations in the PSF, we drew stellar selection boxes around these branches in the $g$-band images, with a consistent faint-end magnitude cut of $g < 24$ mags to limit contamination. The upper limit on the radius was determined by considering the sizes of the objects classified as GCs in \cite{Strader2011}'s kinematic sample and ensuring that all of these, except for a few extreme outliers, survived the cull. An example of this selection is shown in Figure~\ref{fig:GCsel1} for the +0+0 field. Our selection is deliberately conservative, since non-GC contamination is dealt with in the subsequent analysis.

\item \textit{Colour selection}: Because of the bimodal nature of M87's GCs, they lie on two distinct branches in colour-magnitude space, as can be seen in Figure~\ref{fig:GCsel2}. The diagram has an overdensity at $g-i\sim0.4$ due to main-sequence turn-off (MSTO) stars in the intervening Milky Way halo and, on the red end at $g-i\sim2-2.5$, a contribution from nearby, low-luminosity disk stars, while unresolved background galaxies span colour space at the faintest magnitudes. This leaves two distinct peaks at $g-i\sim0.7$ and $g-i\sim1.0$ which we identify as the blue and red GC populations, though we note that the peak colour of the interloping Sagittarius stream also lies close to the blue GC population (although these stars are typically brighter than GCs). We therefore isolate the GC branches by imposing the cuts: 

\noindent
\centerline {$0.2 < g-r < 1.0$} \\
\centerline {$0.5 < g-i < 1.45$} \\
\centerline{$i>18.0$.}
\end{enumerate}
These final cuts leave a catalogue of 17620 GC candidates spanning radii from 1.3 kpc to 445 kpc, though the azimuthal coverage is only complete out to 240 kpc. We therefore define a second catalogue of 10784 objects which extend out to this radius. This is the catalogue which we use in the analysis that follows. A sample of the full catalogue is presented in Table~\ref{tab:cat}, and the full version is available online. The photometry provided in the catalogue has not been corrected for dust extinction, though we do correct for this in our analysis, using the dust maps of \cite{Schlafly2011}.

\begin{table*}
\small
 \centering

\begin{tabular}{p{1.2cm}p{1.3cm}p{1.2cm}p{1.2cm}p{1.2cm}p{1.2cm}p{1.2cm}p{1.2cm}p{1.2cm}p{1.2cm}p{1.2cm}p{0.5cm}}\hline    
  RA (deg) & Dec (deg) & $u$ (mag) & $g$ (mag) & $r$ (mag) & $i$ (mag) & $z$ (mag) & p(red) & p(blue) & p(MW) & p(int) & flag \\\hline
187.80388 & 11.48634 & 24.59$\pm$0.09 & 23.93$\pm$0.03 & 23.62$\pm$0.05 & 23.25$\pm$0.05 & 23.25$\pm$0.14 & 0.00$\pm$0.00 & 0.44$\pm$0.02 & 0.00$\pm$0.00 & 0.56$\pm$0.02 & 0 \\
188.03903 & 11.48655 & 25.51$\pm$0.21 & 23.99$\pm$0.03 & 23.48$\pm$0.04 & 23.18$\pm$0.05 & 23.26$\pm$0.14 & 0.22$\pm$0.03 & 0.06$\pm$0.03 & 0.00$\pm$0.00 & 0.71$\pm$0.02 & 0 \\
188.01770 & 11.48819 & 24.21$\pm$0.07 & 22.30$\pm$0.02 & 21.56$\pm$0.02 & 21.20$\pm$0.02 & 20.91$\pm$0.03 & 0.00$\pm$0.00 & 0.00$\pm$0.00 & 0.00$\pm$0.00 & 1.00$\pm$0.00 & 0 \\
187.62814 & 11.48812 & 24.45$\pm$0.08 & 23.80$\pm$0.03 & 23.11$\pm$0.03 & 22.97$\pm$0.04 & 22.75$\pm$0.09 & 0.03$\pm$0.02 & 0.00$\pm$0.00 & 0.00$\pm$0.00 & 0.97$\pm$0.02 & 0 \\
187.75686 & 11.48803 & 24.49$\pm$0.09 & 23.44$\pm$0.03 & 22.65$\pm$0.03 & 22.47$\pm$0.03 & 21.81$\pm$0.04 & 0.00$\pm$0.00 & 0.00$\pm$0.00 & 0.00$\pm$0.00 & 1.00$\pm$0.00 & 0 \\
188.38547 & 11.48904 & 25.11$\pm$0.14 & 22.99$\pm$0.02 & 22.21$\pm$0.02 & 21.77$\pm$0.02 & 21.52$\pm$0.04 & 0.00$\pm$0.00 & 0.00$\pm$0.00 & 0.00$\pm$0.00 & 1.00$\pm$0.00 & 0 \\
188.12481 & 11.48844 & 22.81$\pm$0.03 & 21.75$\pm$0.02 & 21.30$\pm$0.02 & 21.10$\pm$0.02 & 20.94$\pm$0.03 & 0.00$\pm$0.00 & 0.94$\pm$0.01 & 0.00$\pm$0.00 & 0.06$\pm$0.01 & 0 \\
187.86881 & 11.48895 & 24.28$\pm$0.07 & 23.27$\pm$0.02 & 22.90$\pm$0.03 & 22.71$\pm$0.03 & 22.71$\pm$0.09 & 0.00$\pm$0.00 & 0.78$\pm$0.03 & 0.00$\pm$0.00 & 0.22$\pm$0.03 & 0 \\
188.34479 & 11.49241 & 23.22$\pm$0.03 & 21.27$\pm$0.02 & 20.37$\pm$0.02 & 20.02$\pm$0.02 & 19.75$\pm$0.02 & 0.00$\pm$0.00 & 0.00$\pm$0.00 & 0.61$\pm$0.01 & 0.39$\pm$0.01 & 0 \\
188.45943 & 11.49544 & 24.65$\pm$0.07 & 23.84$\pm$0.03 & 23.44$\pm$0.04 & 23.33$\pm$0.05 & 23.01$\pm$0.12 & 0.00$\pm$0.00 & 0.24$\pm$0.01 & 0.65$\pm$0.01 & 0.11$\pm$0.00 & 0 \\\hline
\end{tabular}
\caption{Catalogue data. Magnitudes are PSF-fitted using SExtractor and PSFex, and are measured in the CFHT/MegaPrime filter system without correcting for dust extinction. Columns 1 and 2 list the RA and Dec of the sources; columns 3 - 7 present the \textit{ugriz} magnitudes with associated uncertainties; columns 8 - 11 give the the probabilities of belonging to each of the four populations with associated uncertainties, and column 12 gives the object's velocity if it is included (and classified as a GC) in the kinematic catalogue of \protect\cite{Strader2011} and is set to zero otherwise. The full version of this table is available in machine-readable form in the online journal.}
\label{tab:cat}
\end{table*}

\section{GC Populations}
\label{sec:pops}
The GCs in M87 comprise two or three separate populations with distinct colours, spatial distributions, GC luminosity functions (GCLFs) and formation histories, the standard scenario being a two-component model with a redder, more compact (and usually referred to as metal-rich) population existing alongside a bluer, more extended one \citep[e.g.][]{Tamura2006b}. These differences act as filters to help in picking out the GC populations from the interloping objects. We therefore model the catalogue as being composed of four distinct components, including the red and the blue GCs as well as two contaminant populations. We allow for one population of interloping stars from the Milky Way disk and halo and the Sagittarius stream, whose colour distributions we can model in detail, and a second interloping population of uniform colour, which could include distant background galaxies and stars from other sources (e.g., the Virgo overdensity). Within this paradigm, each GC population has an azimuthally uniform distribution with a S\'ersic profile in radius, a Gaussian distribution in $g-r$ and $g-i$ colours and a Gaussian GCLF in the $g$-band. This form for the GCLF was chosen to facilitate comparison with previous studies \citep[e.g. ][]{Hanes1977, Tamura2006a, Harris2014}. Likewise, we chose to use S\'ersic profiles in radius following previous authors \citep{Strader2011, Agnello2014}. We also allow for radial gradients within the colour distributions of each GC population, using the functional form
\begin{equation}
 \begin{split}
  \mu' &= \mu - \nu \log(R/R_{eff}) \\
 \end{split}
\label{eq:radgrad}
\end{equation}
where $\mu$ and $\nu$ define a log-linear relation between radius and the peak $\mu'$ of the Gaussian and are inferred for each colour distribution, with $R_{eff}$ set equal to 16 kpc, the effective radius of the starlight as reported in \cite{Kormendy2009}. 

\begin{figure}
  \centering
  \includegraphics[trim=30 0 10 0,clip,scale=0.45]{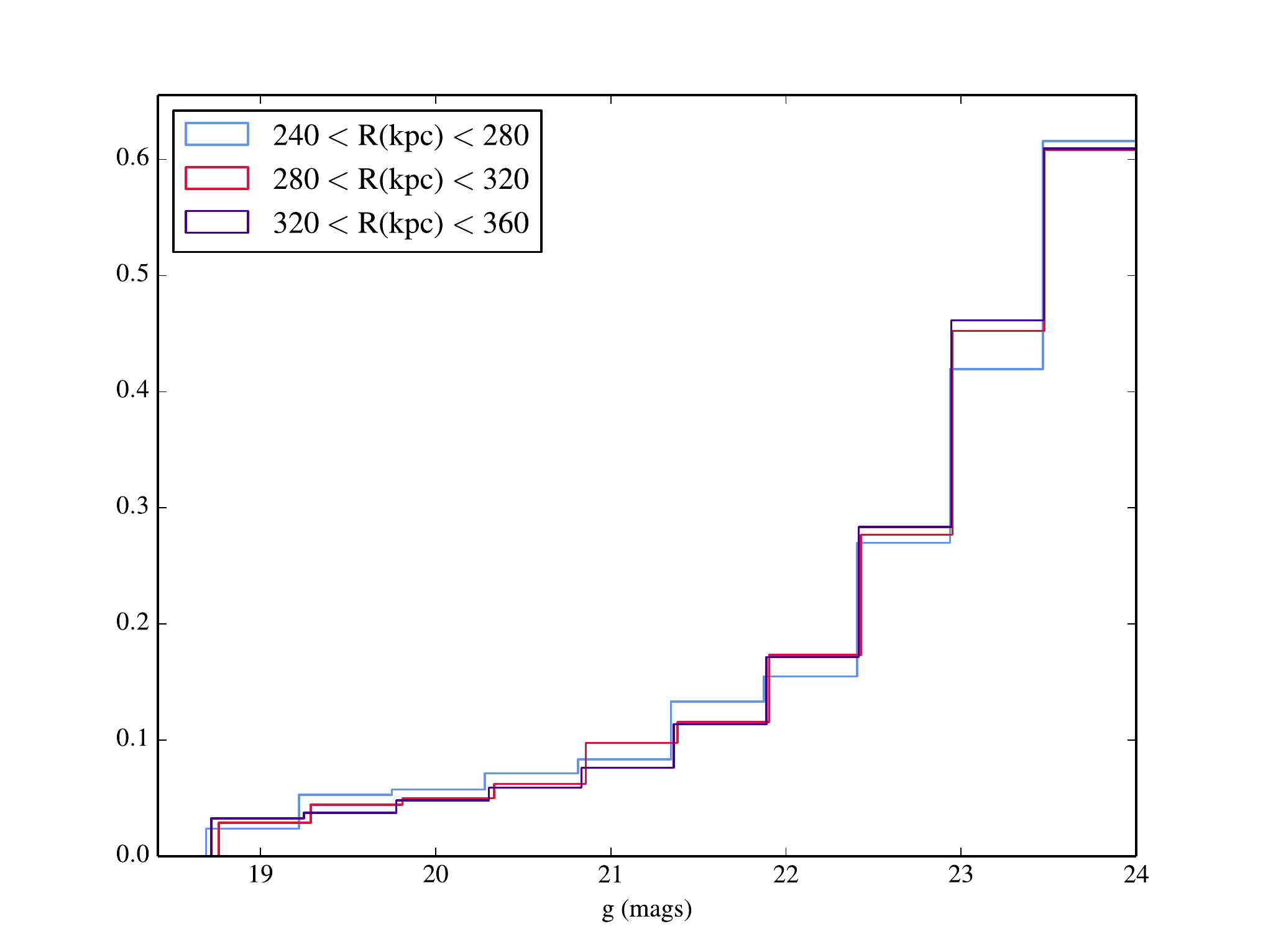}
\caption{We based our model for the LF of the uniform-colour interloping populations on the LF of the full GC candidate catalogue at large radii. The assumption is that the GC profiles have largely died away by the time we reach these radii, meaning we can attribute the LF here entirely to the interlopers: otherwise, using this as our LF could artificially suppress the sizes of the GC populations in our inference. These histograms show the LF in the outermost bins, and we see that the evolution is minimal.}
\label{fig:LF}
\end{figure}

The uniform-colour interloping population is also uniform in space, and has a luminosity function (LF) based on the form of the catalogue's LF at large radii, where the GC profiles are assumed to have largely died away. To check this was a reasonable assumption, we binned the full catalogue radially and compared how its LF changed as a function of radius in the outermost bins. As Figure~\ref{fig:LF} shows, there is negligible variation at large radii. Finally, the Milky Way interlopers have a uniform spatial distribution -- as we are only sampling small fractions of the Sagittarius/Milky Way systems in our field of view -- and a colour distribution based on a combination of synthetic survey data for the Milky Way \citep[using the code Galaxia, ][]{Sharma2011} and for Sagittarius using models for the star formation history from observations in the SDSS Stripe 82 \citep{deBoer2015}. While the radial and colour distributions are the strongest diagnostics here, the relatively faint magnitude limit of our catalogue makes the LF a useful additional tool for deselecting contaminants, whose densities are expected to increase rapidly at the faint end. 

We infer the parameters of our model using Bayes theorem. This states that, given data $\vec{D_k}$ and model parameters $\vec{M}$, the posterior distribution for the model is given by the product
\begin{equation}
 P(\vec{M}|\vec{D_k}) \propto P(\vec{D_k}|\vec{M})~ P(\vec{M})
\end{equation}
where the first term on the right represents the likelihood of the data $\vec{D_k}$ given the model and the second term is the prior, which we assign to the model based on our existing knowledge. In our model, the data $\vec{D_k}$ comprises a vector containing the galactocentric radius, the $g$-band magnitude and the $g-i$, $g-r$ colours,
\begin{equation}
 \vec{D_k} =
\left(
\begin{array}{c}
R_k \\
g_k \\
(g-i)_k \\
(g-r)_k 
\end{array},
\right)
\end{equation}
and the model parameters $\vec{M}$ are listed in Table~\ref{tab:params}. Assuming flat priors on all parameters, we can absorb these into our proportionality constant to write the posterior as
\begin{equation}
 P(\vec{M}|\vec{D_k}) \propto P(\vec{D_k}|\vec{M}),
\end{equation}
and, as each catalogue entry constitutes an independent observation, we can write the joint posterior density function (pdf) for the $k$ objects in the catalogue as the product
\begin{equation}
P(\vec{M}|\vec{D}) = \prod_{k} P(\vec{M}|\vec{D_k})
\label{eq:six}
\end{equation}
where $\vec{D}$ now represents the full dataset. 

With this formalism established, we now turn to the detailed form of the likelihood function $P(\vec{D_k}|\vec{M})$ of observing a single object given a set of model parameters. As explained at the beginning of this Section, our model comprises four distinct components; as such, the likelihood function for the $k^{th}$ object is simply their weighted sum,
\begin{equation}
P(\vec{D_k}|\vec{M}) = (f_r P_r + f_b P_b + f_s P_s + f_i P_i) \frac{A_{eff}}{A},
\label{eq:gc}
\end{equation}
where the subscripts $r$, $b$, $s$ and $i$ correspond to the red GC, blue GC, interloping stellar and unclassified interloper populations respectively, and the $f$ coefficients represent their relative fractions, normalised such that
\begin{equation}
f_r + f_b + f_s + f_i = 1
\end{equation}
within the area covered by our data, out to 240~kpc. The factor $\frac{A_{eff}}{A}$ is the fractional effective area of our detection image at that object's galactocentric radius, accounting for the masking of bad pixels and bright objects, as explained in Section~\ref{sec:phot}. In the log-likelihood calculation, this factor just gives a constant additive term, but it is important for comparing our radial profile with other studies, as it allows us to simply rescale our number count to account for the masked regions.

The likelihoods $P_r$, $P_b$, $P_s$, and $P_i$ for the four populations are then the functions of $R$, $g-i$, $g-r$, and $g$ described at the beginning of this Section, and can be summarised as follows:
\begin{equation}
\begin{split}
P_r = ~&\Sigma(R|R_{e,r},n_r)~\mathcal{G}(\vec{m_r},R| \vec{\mu_r}, \vec{\sigma_r^2}) \\
P_b = ~&\Sigma(R|R_{e,b},n_b)~\mathcal{G}(\vec{m_b},R| \vec{\mu_b}, \vec{\sigma_b^2}) \\
P_s = ~&\mathcal{U}(R)~\mathcal{P}(gi, gr,g) \\
P_i = ~&\mathcal{U}(R,gi, gr) ~\mathcal{P}(g)\\
\end{split}
\end{equation}
for Gaussian distributions $\mathcal{G}$, uniform distributions $\mathcal{U}$, S\'ersic profiles $\Sigma$ and the spline representations of the Milky Way stellar density distribution $\mathcal{P}(gi,gr,g)$ and the uniform interloper luminosity function $\mathcal{P}(g)$. Here, the red and blue GC Gaussians are 4D, with, for instance, the centre of the red distribution given by $\vec{\mu_r} = (\mu_{gi,r}(R), \mu_{gr,r}(R),\mu_{g,r})$ with the radial dependence of the colours explicitly noted. Note that the GCLF Gaussian is explicitly truncated at $g = 24$ to account for our (imposed) selection function, which is assumed to be flat otherwise.

\begin{figure*}
\centering
\includegraphics[width=\textwidth]{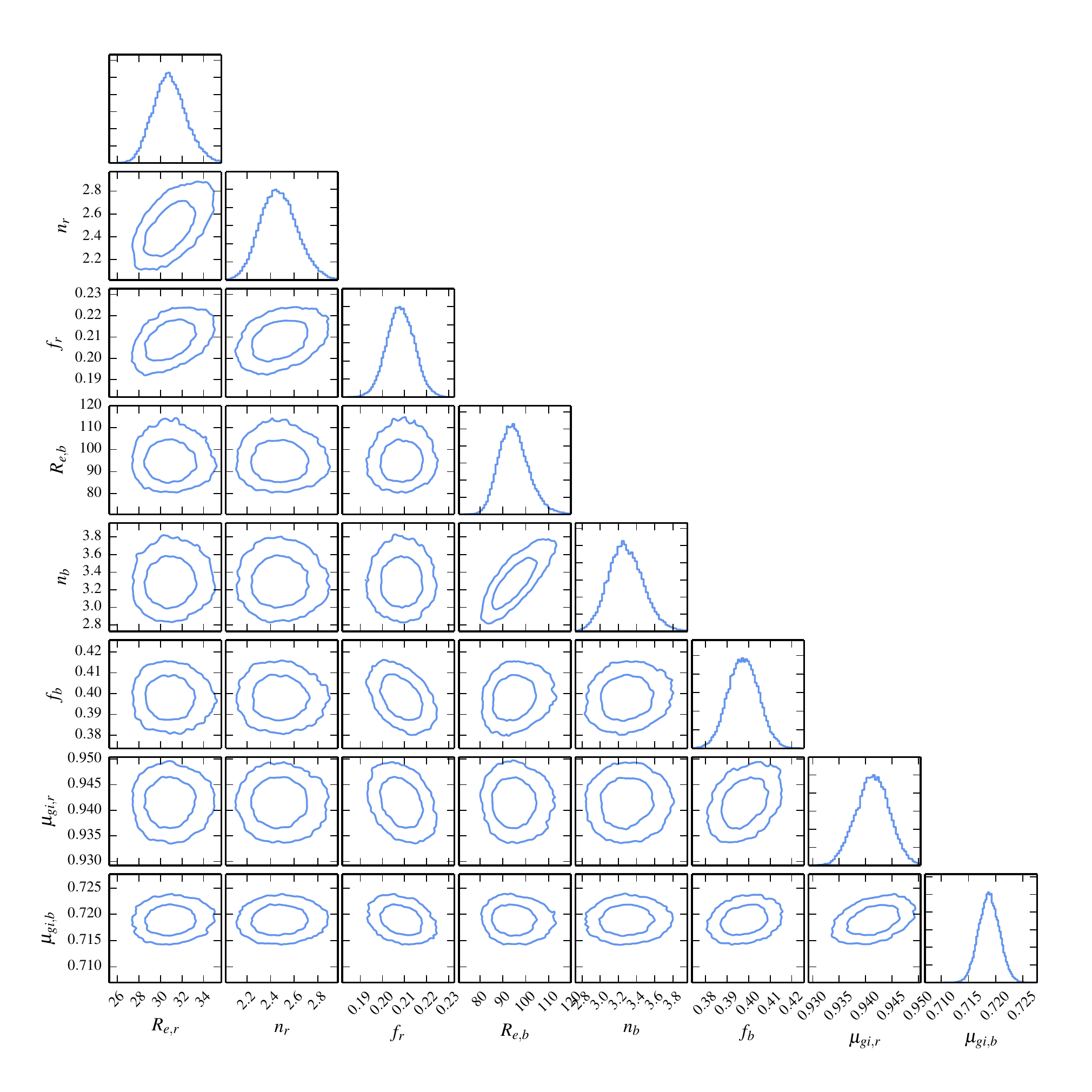}\hfill
\label{fig:tri}
\caption{Inference on the model parameters. Our 23-parameter space includes S\'ersic radial profiles and Gaussian colour distributions for the red and blue GC populations, plus separate components of interloping stars and galaxies. Here we show the posterior marginalised over a number of parameters, in order to emphasise those dictating the GC profiles, with contours representing the $68^{\text{th}}$ and $95^{\text{th}}$ percentiles.}
\end{figure*}

The final model then has 23 free parameters, which we explore using the ensemble-sampling code \texttt{emcee} \citep{Foreman-Mackey2013}. Our inferred posteriors are shown in Figure~\ref{fig:tri}, with their maximum-likelihood values and associated uncertainties listed in Table~\ref{tab:params}.

\begin{figure*}
\centering
	\subfigure{\includegraphics[trim=20 0 20 0,clip,width=0.48\textwidth]{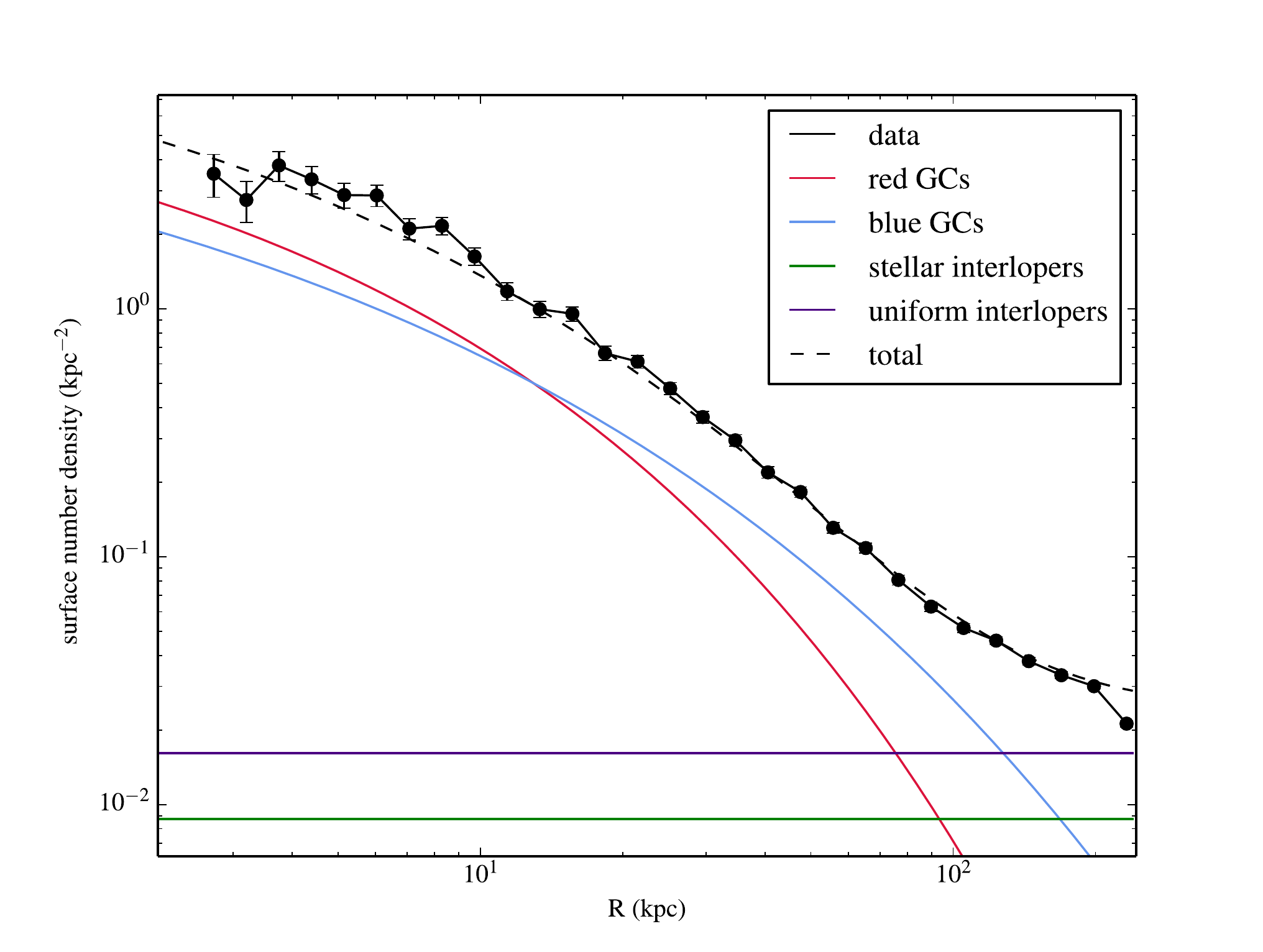}}\hfill
  	\subfigure{\includegraphics[trim=20 0 20 0,clip,width=0.48\textwidth]{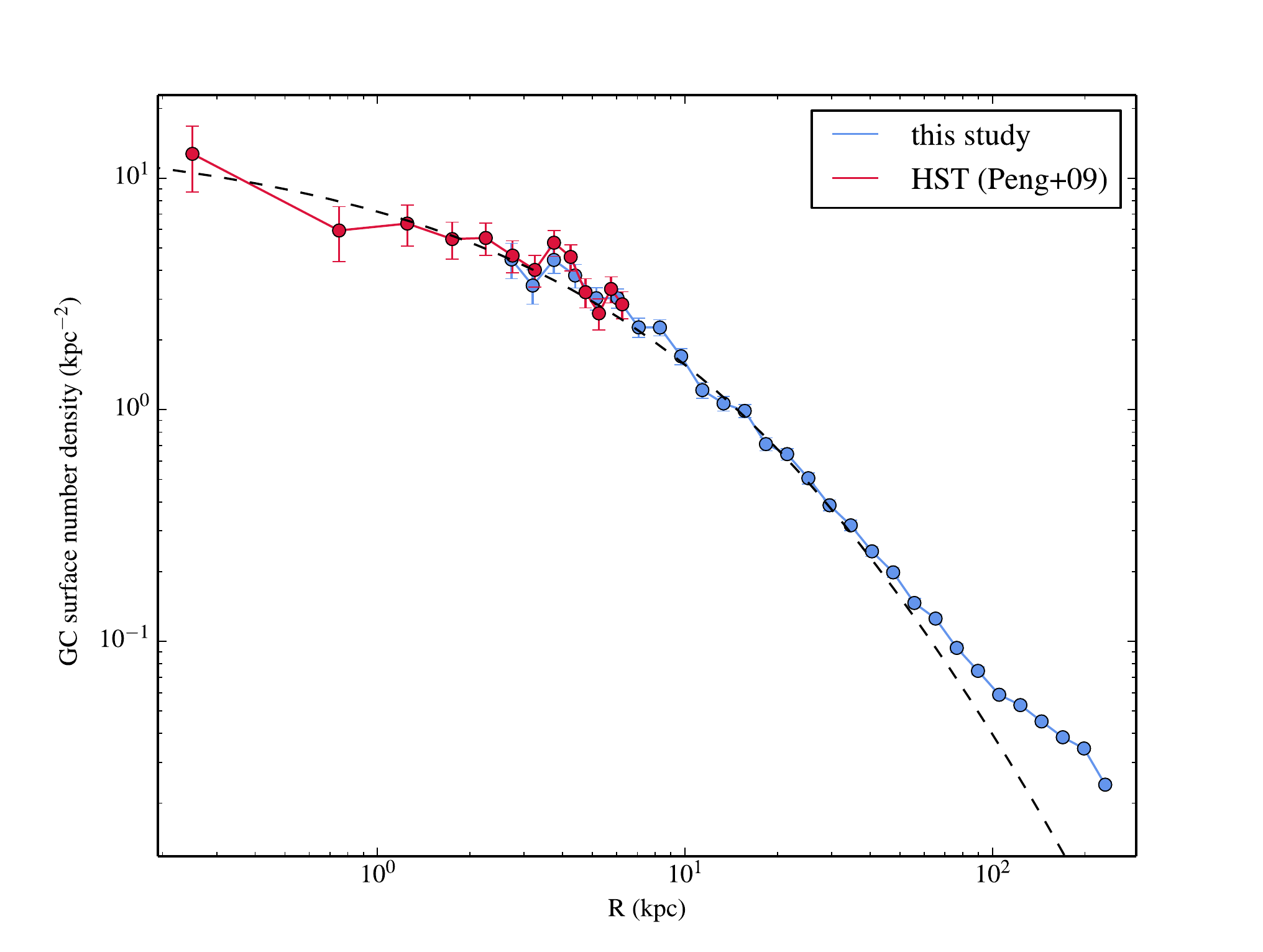}}\hfill
\label{fig:R}
\caption{Left: Inferred surface density. The contributions from the red GC, blue GC and interloping components are plotted, together with the total inferred surface density, shown by the black dashed line. The data are overplotted in circles. 
Right: Comparison with the HST catalogue of \citet{Peng2009}. These data, which extend to smaller radii than our catalogue, continue to follow our inferred GC profile. Note that the surface density of objects in our catalogue deviates from the GC profile at large radii because of the increasing relative importance of the interloper population; conversely, the dominance of the GC populations at the smallest radii means our data can be well-approximated by the GC curves alone in this region. }
\end{figure*}

\begin{table*}
\centering
\begin{tabular}{|c|ccc|}\hline
\multirow{10}{*}{Red GCs} &   &  radial gradient model & fixed-Gaussian model \\\cline{1-4}
    & $R_e$ (kpc) & $30.6 \pm 1.6$ & $24.1 \pm 0.9$  \\
    & $n$ & $2.41 \pm 0.16 $ & $2.012 \pm 0.12$  \\
    & $\mu_{gi}$ (mags) & $0.944 \pm 0.003$ & $0.919 \pm 0.003$ \\
    & $\sigma_{gi}$ (mags) & $0.084 \pm 0.002$ & $0.100 \pm 0.002$  \\
    & $\mu_{gr}$ (mags) & $0.597 \pm 0.002$  & $0.588 \pm 0.002$   \\
    & $\sigma_{gr}$ (mags) & $0.045 \pm 0.002$ &  $0.055 \pm 0.002$  \\
    & $f_r$ & $0.204 \pm 0.009$ & $0.217 \pm 0.009$  \\
    & $\mu_g$ (mags) & $24.60 \pm 0.38$ & $24.68 \pm 0.29$  \\
    & $\sigma_{g}$ (mags) & $1.65 \pm 0.13$ & $1.72 \pm 0.10$   \\
    & $\nu_{gi}$ (mags) & $0.14 \pm 0.01$ &  \\
    & $\nu_{gr}$ (mags) & $0.038 \pm 0.003$ & \\\hline
\multirow{9}{*}{Blue GCs} & $R_e$ (kpc) & $95.0 \pm 6.8$ & $109.0 \pm 8.5$ \\
    & $n$ & $3.23 \pm 0.2$ & $2.74 \pm 0.19$  \\
    & $\mu_{gi}$ (mags)& $0.719 \pm 0.002$ & $0.677 \pm 0.001$  \\
    & $\sigma_{gi}$ (mags) & $0.046 \pm 0.001$ & $0.051 \pm 0.001$  \\
    & $\mu_{gr}$ (mags)& $0.459 \pm 0.002$ & $0.447 \pm 0.002$  \\
    & $\sigma_{gr}$ (mags) & $0.034 \pm 0.002$ &  $0.036 \pm 0.002$  \\
    & $f_b$ & $0.398 \pm 0.007$ & $0.410 \pm 0.007$  \\
    & $\mu_g$ (mags) & $23.65 \pm 0.10 $ & $23.77 \pm 0.14$  \\
    & $\sigma_{g}$ (mags)& $1.37 \pm 0.05$ & $1.42 \pm 0.06$  \\
    & $\nu_{gi}$ (mags)& $ 0.080 \pm 0.003$ &  \\
    & $\nu_{gr}$ (mags) & $0.023 \pm 0.003$ &  \\\hline
\multirow{2}{*}{Interlopers} & $f_s$ & $0.134 \pm 0.005$ & $0.134 \pm 0.005$  \\
    & $f_i$ & $0.264 \pm 0.006$ & $0.239 \pm 0.006$  \\\hline
\end{tabular}
\caption{Inferred parameters, for both fixed colour Gaussians and colour distributions whose mean decreases as a function of radius. Listed are the maximum-likelihood values from the pdf, along with uncertainties given by their 16th and 84th percentiles. All parameters were assigned uniform priors.}
\label{tab:params}
\end{table*}

\section{Discussion}
Our inferred surface density profile is shown in Figure~\ref{fig:R}, and the parameters of the fit can be found in Table~\ref{tab:params}. In agreement with previous studies \citep[e.g. ][]{Cote2001, Durrell2014}, we find that the blue population within our catalogue is significantly more populous than the red, with the total GC count made up of $\sim 1/3$ red GCs and $2/3$ blue out to our cut-off radius of 240 kpc and down to 24th magnitude in the $g$-band. The blue population is also much more extended. However, given that the LF of the red population is roughly one magnitude fainter than that of the blue, we expect the red population to be larger overall: assuming our inferred distributions remain valid at large radii and faint magnitudes, we find that there should be $\sim6030$ (mostly faint) red GCs and $\sim5300$ blue GCs, giving  $11330_{-300}^{+1500}$ GCs in total. This is consistent with most previous estimates \citep[e.g.][]{Tamura2006a}, though it is smaller than the number count of $N = 14,520 \pm 1190$ reported by \cite{Durrell2014}. That study was focussed on a cluster-wide GC survey and did not include the contaminant model for Sagittarius and Milky Way stars that we implement here, so it is possible that their blue GC population may include a contribution from Sagittarius stars, whose colour-magnitude distribution is very similar to that of the blue population. Indeed, their number counts at large radii are slightly larger than what our S\' ersic profiles predict. On the other hand, given that their data extend further out than the catalogue we use here, it is also possible that the blue population could shift to a different profile at these large radii. If this could be demonstrated, this would be an interesting result which could provide strong evidence for GC accretion. However, while \cite{Durrell2014} note that the mean slope of the blue GC surface density appears to become shallower at $R \sim 250$ kpc, they also stress that the uncertainties at these radii are too large to facilitate any strong conclusion.  

Previous studies of GC systems in elliptical galaxies have shown that the red GC distribution tends to follow the starlight \citep[e.g.][]{Geisler1996}, and this has been verified for M87 through studies such as that of \citet{Durrell2014}. We test this for our model by comparing our S\' ersic profile for the red GCs with a fit to the surface brightness distribution of the starlight given in \citet{Kormendy2009}, and find that the two profiles do indeed match closely at small and intermediate radii ($< 100$ kpc), though at larger radii, the red GCs die away more rapidly. As M87 is known to have an extended stellar envelope, possibly built up through merger events \citep[e.g.][]{Hausman1978}, and virtually all GC formation scenarios have the red population forming \textit{in situ} \citep[e.g.][]{Ashman1992, Cote1998}, it is perhaps not surprising that this GC population does not trace the starlight at these larger radii. We also note that, in spite of the increased depth of our catalogue, our average contaminant surface density of 0.58 arcmin$^{-2}$ is similar to that found in eg. \cite{Strader2011} and \cite{Tamura2006a}, and that this increased depth is valuable in terms of GC numbers; according to our model, we expect the GC populations together to constitute $\sim$58\% of the objects with $g\sim24$ out to 240 kpc. We compare the surface density of GCs as seen in the HST data of \cite{Peng2009} with our S\'ersic profiles -- shown in the right panel of Figure~\ref{fig:R} -- and find that our model is also a good fit at the innermost radii, where our catalogue does not reach. This is encouraging, and implies that any tidal distruption that may have occurred cannot have been effective in removing the central GC population.

We find luminosity functions for the two GC components whose peaks are comparable with the limiting magnitude of the catalogue, with $\mu_{g,r} = 24.60 \pm 0.38$ and $\mu_{g,b} = 23.65 \pm 0.10$ magnitudes, which are consistent with previous results. \cite{Tamura2006a} modelled the $V$-band distribution of the total GC population using data extending out to $\sim 0.5$ Mpc and found a turnover magnitude of $V = 23.62 \pm 0.06$ and $\sigma_V = 1.40 \pm 0.04$, while \cite{McLaughlin1994} found a peak at $V = 24.2$ mags with a similar width of $\sigma_V = 1.73$. The GCLF is also consistent with that inferred by \cite{Peng2009} for GCs at smaller radii: they find a turnover magnitude for the total population $I = 22.53 \pm 0.05$, and when we cross-correlate the catalogues to calculate a colour correction, this translates to a $g$-band magnitude $g \sim 23.75$. Adopting the \citet{Tamura2006a} measurement of the total $V$-band galaxy luminosity of $M_V = -22.46$, we calculate the specific frequency $S_N$ of M87's GC system as $11.76 \pm 2.1$, where we have included an uncertainty of $0.1$ mags on the absolute magnitude $M_V$. As noted in other studies \citep[e.g.][]{McLaughlin1994}, this is extremely high -- the average $S_N$ of other ETGs in Virgo is $\sim 5$ \citep{Harris1991} and the cluster $S_N$ is $\sim 3$ \citep{Durrell2014} -- and could be related M87's location at the centre of a massive cluster. For instance, as the BCG in Virgo, M87 may have undergone an unusually large number of GC-triggering mergers \citep[e.g.,][]{Ashman1992}, or alternatively it may have accreted many GCs from smaller satellite galaxies \citep[e.g.,][]{Forte1982}. Support for these environment-focussed interpretations comes from the fact that a number of other BCGs have been found to have similarly high specific frequencies \citep[e.g.,][]{McLaughlin1994}. 

We can also use the blue GC population as an independent distance indicator by comparing the peak brightness of the GCLF with that of other metal-poor GC systems. We take the turnover magnitude of these to be
MV = −7.66 $\pm$ 0.09 \citep{DiCriscienzo2006} and use the \citet{Peng2009} HST catalogue to determine an empirical $g-V$ correction of $g-V$ = 0.323 $\pm$ 0.049. This gives a distance modulus $DM = 30.99 \pm 0.14$, or $D_{\rm L} = 15.79 \pm 1.04$ Mpc, in very good agreement with other measurements of M87's distance modulus \citep[e.g.,][]{Ferrarese2000,Mei2007}.

M87's GC population as seen in CFHT has previously been modelled by \cite{Strader2011}, where the full photometric sample from CFHT was used to characterise the underlying distribution of a smaller kinematic subsample. This subsample was later used by \cite{Agnello2014} in a series of dynamical models, where they determined the apparent spatial distribution of the \textit{kinematic} sample. The very significant difference between our S\'ersic profiles and those in \citet{Agnello2014} -- compare their $R_{e,r} = 6$ kpc and $R_{e,b} = 190$ kpc with our results in Table~\ref{tab:params} -- can be understood by the fact that the spectroscopic subsample incorporates some non-trivial selection function, meaning the radial distributions are indeed different. Our S\'ersic profiles are similar to those inferred in \cite{Strader2011}, though we find a S\'ersic index for the red population which is significantly smaller than the $n_r=5.33$ fitted in that work. It is important to note that our GC populations differ from theirs slightly in that ours extend to a different magnitude limit \citep[applied a magnitude limit of $i<23$ mags in the SDSS i-band, which has only a small offset from the MegaCam i-band magnitude]{Strader2011} and smaller galactocentric radii \citep[in][they fitted the data in radial bins starting at $\sim$ 5 kpc]{Strader2011}; this could contribute in part to the difference, which in our models is driven by the objects at small radii, where the red GCs are most populous. Further, as noted in that paper, it is hard to assess the uncertainties on the S\'ersic parameters due to their strong correlations.

\subsection{Importance of correctly modelling the interlopers}
An additional feature of our study is the more accurate and physically-motivated model that we use for the interloper colour distribution. While the standard way to treat background contamination in these systems in the past has been to distribute them uniformly in both space and colour, the latter assumption is not justified for the case of intervening Milky Way and stream stars, which we should expect to have colours strongly clustered around the MSTO point and the red-giant branch (RGB) with a scatter determined by the spread in stellar age. This gives a particularly strong bias in the case of Sagittarius, whose RGB lies virtually on top on the blue GC distribution in colour space. In our model, then, what a uniform-colour contaminant model would assign to the blue population at large radius, we might be more likely to classify as an interloping star. 

To test the sensitivity of the inference to different interloper distributions, we performed the inference with the following models: 
\begin{enumerate}
 \item a single interloper population, uniform in both space and colour
 \item a single population of interloping (Milky Way and Sagittarius) stars
 \item two populations, but with the contribution from Sagittarius excluded from the stellar model
 \item a single population of Milky Way stars.
\end{enumerate}

We found that case (i), assuming only uniformly distributed contaminants, required the scale radius and S\'ersic index of the blue GC population to be large: $R_{e,b} \sim$ 250 kpc, $n \sim$ 3.5. As explained above, this is due to the overlap of the blue GCs and the Sagittarius stars in the CMD. Case (ii), requiring all the interlopers to obey the Milky Way/Sagittarius model, converged on a solution in which Sagittarius dominated the catalogue to give $R_{e,b} \sim$ 30 kpc, clearly highlighting the importance of including both sources of contamination. In case (iii), the lack of Sagittarius resulted in the stellar interloper distribution being a poor description of the data, such that the stellar fraction $f_s \to 0$ and the S\'ersic profiles of case (i) were returned. Finally, case (iv) recovered profiles comparable to those obtained from the Sagittarius+Milky Way+uniform fits, with the absence of Sagittarius balanced, to some extent, by the absence of the uniform component. However, the fits in colour space were poor due to the mismatch of the Milky Way stars to the underlying contaminant distribution: in particular, the colour model had a bump at the red end due to low-luminosity disk stars, and this was obviously discrepant with the data. This, together with the significant $f_s$ component that we infer, shows that the classification of objects as GCs versus contaminants is strongly dependent on our choice of contaminant model. Further, as the Sagittarius stellar population is generally bright, the peakiness of its distribution in colour space becomes more pronounced as the depth of the catalogue is reduced, making it even more important in shallower catalogues. Its exclusion could lead to dramatic overestimations of the extent of the blue GC population, as well as how steeply it declines as a function of radius -- which could explain, for instance, the larger total number count found in \cite{Durrell2014}. This is something that must be modelled carefully in order to properly separate the populations.

\subsection{The existence of colour gradients in the GC populations}
\label{sec:grad}
\begin{figure*}
  \centering
  \subfigure{\includegraphics[trim=20 0 20 0,clip,width=0.48\textwidth]{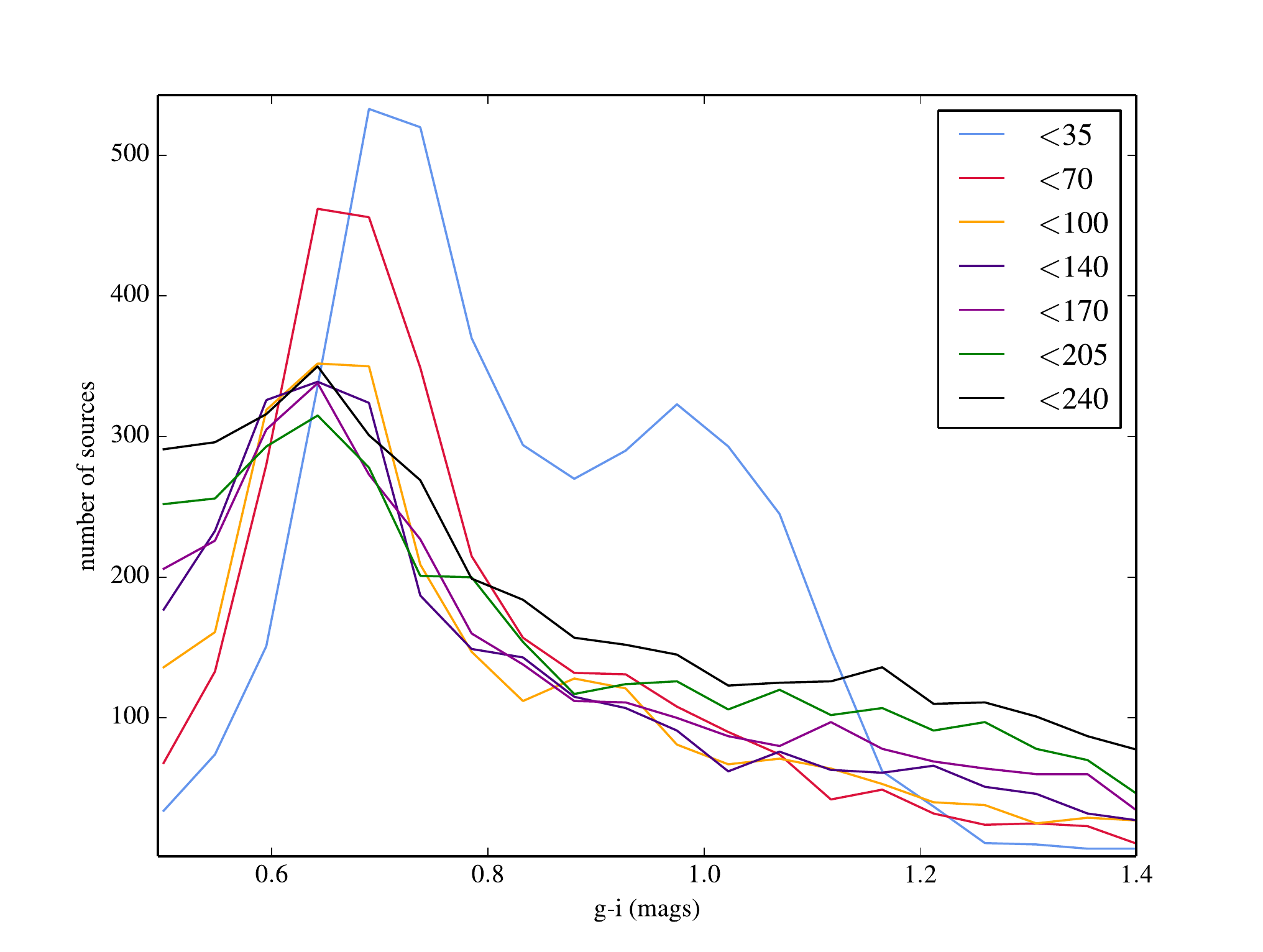}}\hfill
  \subfigure{\includegraphics[trim=20 0 20 0,clip,width=0.48\textwidth]{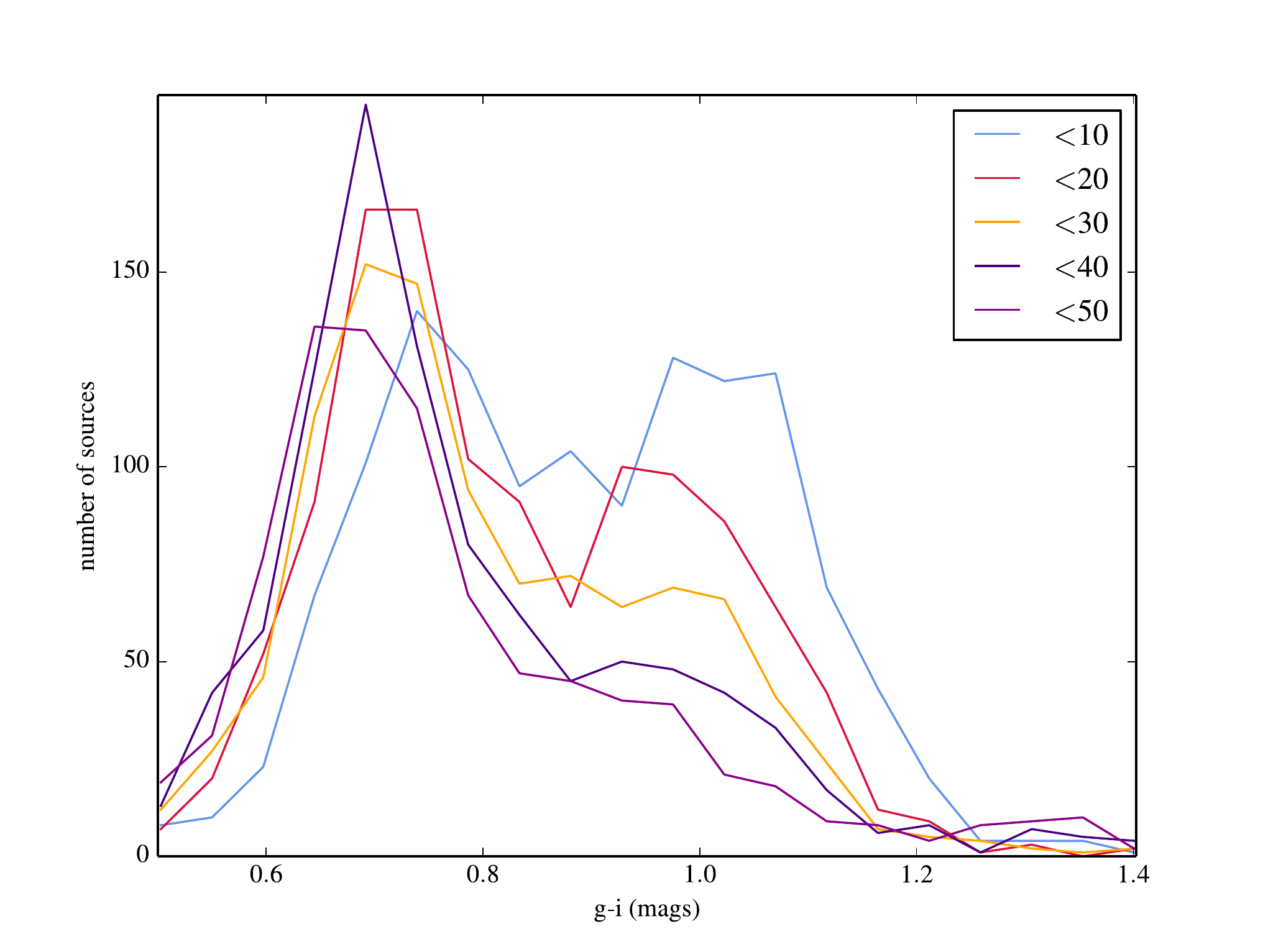}}\hfill
\caption{With increasing galactocentric radius, we see both the red and blue peaks migrating towards bluer colours. The histograms show the colour distributions of the binned data, with the legend denoting the maximum radius of each bin in kpc. In the right panel, we have restricted the dataset to R$<$50 kpc to accentuate the red population.}
\label{fig:tilt}
\end{figure*}

We now examine the cause of the colour structure within each GC population as a function of radius, which was originally noticed in \cite{Strader2006} and recovered in \cite{Harris2009} and \cite{Strader2011}. In those works, the $g-i$ peak within each population was found to shift towards bluer colours as the radius was increased, though while \cite{Harris2009} found this for both the red and the blue GC populations, \cite{Strader2011} only found a significant trend in the red population. In our data, we find a gradient in the colours of both the red and the blue components as a function of galactocentric radius, and plot the colour histograms in radial bins in Figure~\ref{fig:tilt} as a simple way of highlighting this.

In the modelling of Section~\ref{sec:pops}, we interpreted these gradients as existing \textit{within each population}, using Gaussian distributions whose peaks were allowed to vary as a function of radius in a log-linear fashion as in Equation~\ref{eq:radgrad}. This is in line with \cite{Harris2009}, where the observed shift was used to argue for metallicity gradients within each population, which could in turn reflect the enrichment histories of the GCs. However, this is only one interpretation of the shift: \cite{Strader2011} used the same observation to motivate the existence of a third GC population at intermediate colours, with the gradient arising as a result of the different populations merging into one another. Clearly, each of these scenarios has very different implications in terms of the properties and formation histories of the GCs themselves. To distinguish between them, we also performed the inference without allowing for radial gradients, instead forcing the colour distributions to follow Gaussians with fixed peaks (we call this the `fixed-Gaussian' model). We can then compare the inferred parameters for this model to the data, noting that the presence of a significant unmodelled third population would create substantial residuals between the data and the best-fit model in both the colour distribution and the radial profile, whereas colour gradients within the populations would largely affect the colour distribution only.

\begin{figure*}
  \centering
  \subfigure{\includegraphics[trim=20 0 30 0,clip,width=0.48\textwidth]{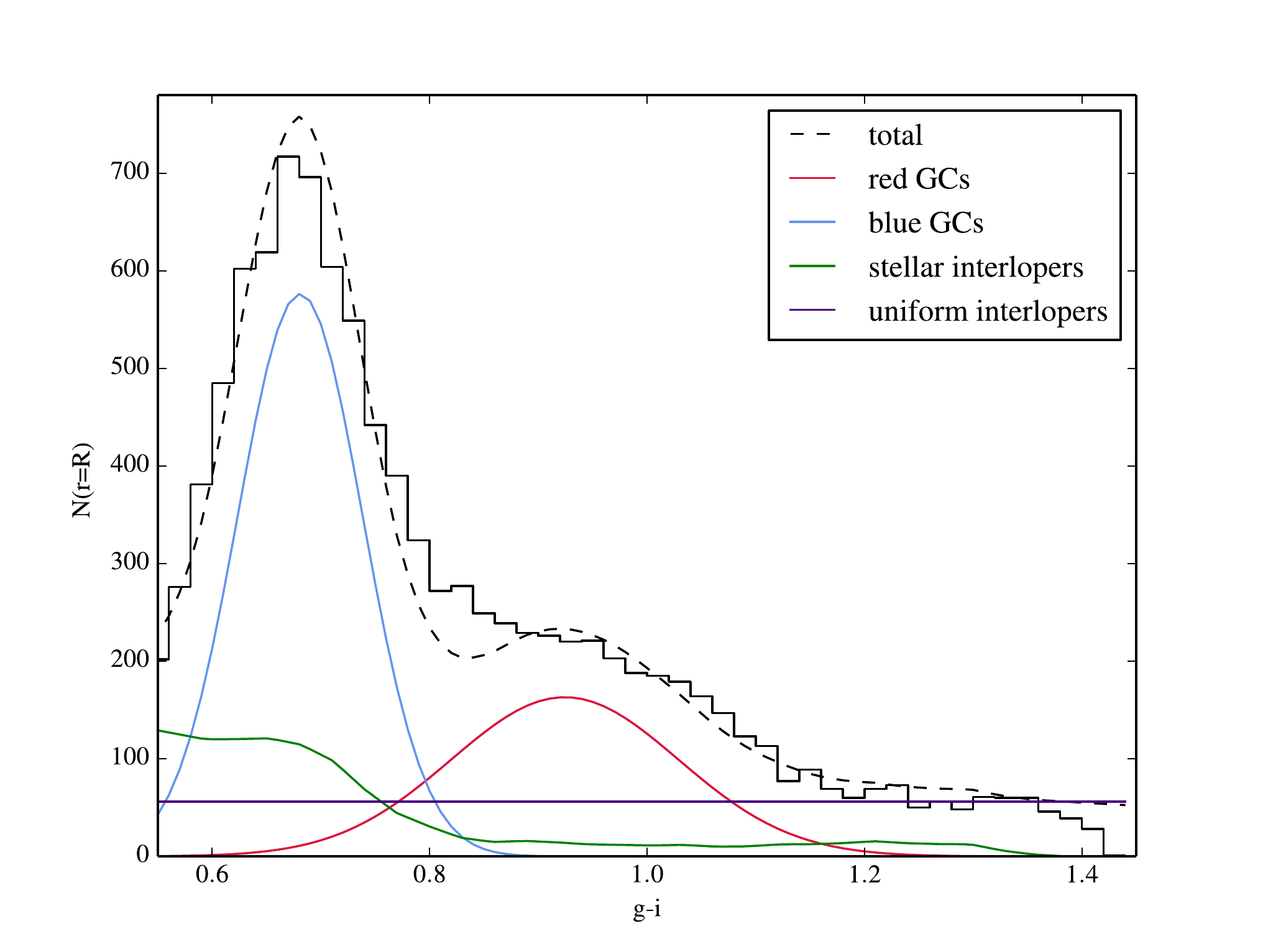}}\hfill 
  \subfigure{\includegraphics[trim=20 0 30 0,clip,width=0.48\textwidth]{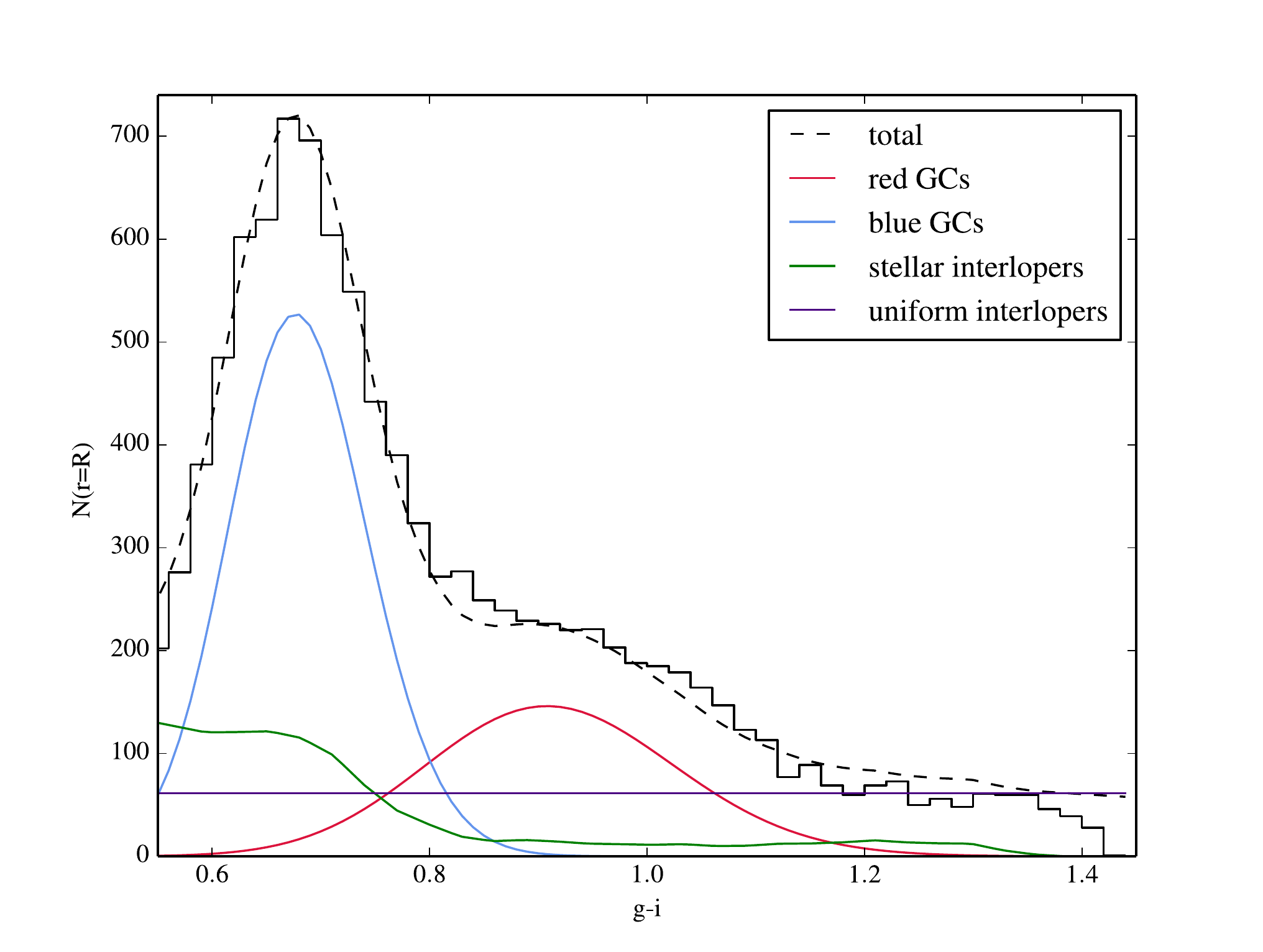}}\hfill     
\caption{Removing the colour bump. Left: the colour distribution of the catalogue (shown by the histogram) is discrepant with the inference from a fixed-peak Gaussian model (shown by the dashed line) at intermediate colours.This motivates a model in which the colour distributions are also a function of radius. Right: allowing for radial gradients in the colour distributions of the GCs alleviates the discrepancy between model and data at intermediate colours.}
\label{fig:colours}
\end{figure*}

Figure~\ref{fig:colours} shows the colour distributions inferred for both the fixed-Gaussian and the radial gradient models. In the former, we see a discrepant bump at intermediate colours which the model cannot recreate.
As explained above, this could be consistent with either colour gradients or a third population. The radial profile inferred for this model, on the other hand, is virtually unchanged from the previous `radial-gradient' model, and does not show any evidence for systematic residuals. This suggests that the GCs are adequately described by a model with only two radial components, but not by two fixed-peak Gaussians in colour, a state of affairs which is more in line with the metallicity-gradient scenario than the three-population hypothesis. A further problem with the three-population scenario is that both our GC populations exhibit colour gradients in the same direction (towards bluer colours). By the reasoning that leads to adding a third population, then, we should also be compelled to add a fourth, and the scenario quickly becomes more complicated.

As discussed in \cite{Harris2009}, the presence of metallicity gradients within the GC populations could have important implications for our understanding of the formation and enrichment histories of the GCs. That is, a population which becomes increasingly enriched at smaller radii, deeper in the gravitational potential well of the galaxy, is a signature of star formation via dissipative collapse, in contrast to that triggered by major mergers and accretion, where material tends to be more spatially mixed. That these gradients appear to exist in both GC populations therefore indicates that both featured some component of dissipative collapse, thus calling into question the importance of formation scenarios in which, for instance, all of the blue GCs are formed in pre-galactic dwarfs and are only later accreted \citep{Cote1998}. Similar relations have been found for other giant ellipticals \citep[e.g.][]{Geisler1996, Forte2001}, and all lend support to this picture of GC enrichment changing as a function of galactocentric radius.

\section{Conclusions}
We have performed a careful subtraction of the M87 stellar light in the CFHT/MegaPrime \textit{ugriz} images in order to extend the wide-field photometry down to small galactocentric radii, and present a catalogue of 17620 GC candidates across a radius range from 1.3 kpc to 445 kpc and to a depth of 24 magnitudes in the $g$-band. By treating the catalogue as being composed of two GC populations and two contaminant populations, we use a Bayesian framework to infer the colour, luminosity and radial profiles of each. 

Our model for the contaminant contribution to the catalogue improves on previous studies by using colours and luminosities that are distributed according to synthetic observations of Milky Way and Sagittarius stars. This is important for distinguishing GCs from stars and correctly characterising their radial profiles, as the red-giant branch of the Sagittarius stream lies extremely close to the blue GC population in colour-magnitude space. The use of this model, in conjunction with a uniformly-distributed component to account for other sources of contamination, significantly changes our inference on the GC distributions, and highlights the importance of modelling the interloper populations in a physically-motivated way. We also confirm previous findings of a colour gradient with galactocentric radius within each GC population, and incorporate this in our model using a log-linear relation.  

The extensiveness and completeness of this catalogue has allowed us to characterise M87's GC populations in a robust, unbiassed way. This makes it an ideal starting point for further studies based on subsamples of these populations, in which selection criteria would otherwise cause the apparent colour and spatial distributions to deviate from the true ones.

\section*{Acknowledgements}
We would like to thank Thomas de Boer for providing the Milky Way and Sagittarius CMDs, and the referee for helpful comments. LJO thanks the Science and Technology Facilities Council (STFC) for the award of a studentship. MWA also acknowledges support from the STFC in the form of an Ernest Rutherford Fellowship.


\begin{thebibliography}{}

\bibitem[\protect\citeauthoryear{Agnello, Evans, Romanowsky \& Brodie}{Agnello
  et~al.}{2014}]{Agnello2014}
Agnello A.,  Evans N.~W.,  Romanowsky A.~J.,    Brodie J.~P.,  2014, \mnras, 442, 3299

\bibitem[\protect\citeauthoryear{Ashman, K.~M., \& Zepf, S.~E.}{Ashman \& Zepf}{1992}]{Ashman1992}
Ashman, K.~M., \& Zepf, S.~E.\ 1992, \apj, 384, 50 


\bibitem[\protect\citeauthoryear{Baum}{Baum}{1955}]{Baum1955}
Baum W.~A.,  1955, Publ. Astron. Soc. Pacific, 67, 328

\bibitem[\protect\citeauthoryear{Bertin \& Arnouts}{Bertin}{1996}]{Bertin1996} 
Bertin, E., \& Arnouts, S.\ 1996, \aaps, 117, 393 


\bibitem[\protect\citeauthoryear{C\^ot\'e, McLaughlin, Hanes, Bridges, Geisler,
  Merritt, Hesser, Harris \& Lee}{C\^ot\'e et~al.}{2001}]{Cote2001}
C\^ot\'e P.,  McLaughlin D.~E.,  Hanes D.~A.,  Bridges T.~J.,  Geisler D.,
  Merritt D.,  Hesser J.~E.,  Harris G. L.~H.,    Lee M.~G.,  2001, \apj, 559, 828

\bibitem[\protect\citeauthoryear{C\^ot\'e, Marzke \& West}{C\^ot\'e
  et~al.}{1998}]{Cote1998}
C\^ot\'e P.,  Marzke R.~O.,    West M.~J.,  1998, \apj, 501, 554


\bibitem[de Boer et al.(2015)]{deBoer2015} de Boer, T.~J.~L., 
Belokurov, V., \& Koposov, S.\ 2015, \mnras, 451, 3489 


\bibitem[\protect\citeauthoryear{Di Criscienzo, M., Caputo, F., Marconi, M., \& Musella, I.}{Di Criscienzo et al.}{2006}]{DiCriscienzo2006}
Di Criscienzo, M., Caputo, F., Marconi, M., \& Musella, I.\ 2006, \mnras, 365, 1357 

\bibitem[\protect\citeauthoryear{Durrell et al.}{Durrell et al.}{2014}]{Durrell2014}
Durrell, P.~R., C{\^o}t{\'e}, P., Peng, E.~W., et al.\ 2014, \apj, 794, 103 


\bibitem[\protect\citeauthoryear{Ferrarese, L. et al.}{Ferrarese et al.}{2000}]{Ferrarese2000}
Ferrarese, L., Mould, J.~R., Kennicutt, R.~C., Jr., et al.\ 2000, \apj, 529, 745


\bibitem[\protect\citeauthoryear{Ferrarese, C\^ot\'e, Cuillandre, Gwyn, Peng,
  MacArthur, Duc, Boselli, Mei, Erben, McConnachie, Durrell, Mihos Christopher,
  Jord\'an, Lan\c{c}on, Puzia, Emsellem, Balogh, Blakeslee, van Waerbeke,
  Gavazzi, Vollmer, Kavelaars, Wo}{Ferrarese et al.}{2012}]{Ferrarese2012}
Ferrarese L., et al. 2012,
  \apjs, 200, 4

\bibitem[\protect\citeauthoryear{Foreman-Mackey, Hogg, Lang \&
  Goodman}{Foreman-Mackey et~al.}{2013}]{Foreman-Mackey2013}
Foreman-Mackey D.,  Hogg D.~W.,  Lang D.,    Goodman J.,  2013, Publ. Astron.
  Soc. Pacific, 125, 306

\bibitem[\protect\citeauthoryear{Forte, Geisler, Ostrov, Piatti \&
  Gieren}{Forte et~al.}{2001}]{Forte2001}
Forte J.~C.,  Geisler D.,  Ostrov P.~G.,  Piatti A.~E.,    Gieren W.,  2001,
  \aj, 121, 1992

\bibitem[\protect\citeauthoryear{Forte, J.~C., Martinez, R.~E., \& Muzzio, J.~C.}{Forte et al.}{1982}]
{Forte1982} Forte, J.~C., Martinez, 
R.~E., \& Muzzio, J.~C.\ 1982, \aj, 87, 1465 

\bibitem[\protect\citeauthoryear{Gebhardt \& Kissler-Patig}{Gebhardt \&
  Kissler-Patig}{1999}]{Gebhardt1999}
Gebhardt K.,  Kissler-Patig M.,  1999, \aj, 118, 1526

\bibitem[\protect\citeauthoryear{Geisler, Lee \& Kim}{Geisler
  et~al.}{1996}]{Geisler1996}
Geisler D.,  Lee M.~G.,    Kim E.,  1996, \aj, 111, 1529

\bibitem[\protect\citeauthoryear{Hanes}{Hanes}{1977}]{Hanes1977}
Hanes D.,  1977, \mnras, 180, 309

\bibitem[\protect\citeauthoryear{Harris}{Harris}{2009}]{Harris2009}
Harris W.~E.,  2009, \apj, 703, 939

\bibitem[\protect\citeauthoryear{Harris}{Harris}{1991}]{Harris1991}
 Harris, W.~E.\ 1991, \araa, 29, 543 

\bibitem[\protect\citeauthoryear{Harris, Morningstar, Gnedin, O'Halloran,
  Blakeslee, Whitmore, Cote, Geisler, Peng, Bailin, Rothberg, Cockcroft \&
  DeGraaff}{Harris et~al.}{2014}]{Harris2014}
Harris W.~E.,  Morningstar W.,  Gnedin O.~Y.,  O'Halloran H.,  Blakeslee J.~P.,
   Whitmore B.~C.,  Cote P.,  Geisler D.,  Peng E.~W.,  Bailin J.,  Rothberg
  B.,  Cockcroft R.,    DeGraaff R.~B.,  2014, \apj, 797,128

\bibitem[\protect\citeauthoryear{Hausman \& Ostriker}{Hausman \& Ostriker}{1978}]{Hausman1978}
Hausman, M.~A., \& Ostriker, J.~P.\ 1978, \apj, 224, 320 


\bibitem[\protect\citeauthoryear{Kormendy, Fisher, Cornell \& Bender}{Kormendy
  et~al.}{2009}]{Kormendy2009}
Kormendy J.,  Fisher D.~B.,  Cornell M.~E.,    Bender R.,  2009, \apj, 182, 216

\bibitem[\protect\citeauthoryear{McLaughlin, Harris \& Hanes}{McLaughlin
  et~al.}{1994}]{McLaughlin1994}
McLaughlin D.~E.,  Harris W.~E.,    Hanes D.~A.,  1994, \apj, 422, 486


\bibitem[Mei et al.(2007)]{Mei2007} Mei, S., Blakeslee, J.~P., 
C{\^o}t{\'e}, P., et al.\ 2007, \apj, 655, 144 


\bibitem[\protect\citeauthoryear{Ostrov, Geisler \& Forte}{Ostrov
  et~al.}{1993}]{Ostrov1993}
Ostrov P.,  Geisler D.,    Forte J.~C.,  1993, \aj, 105, 1762

\bibitem[\protect\citeauthoryear{Peng, Jord\'{a}n, Blakeslee, Mieske,
  C\^{o}t\'{e}, Ferrarese, Harris, Madrid \& Meurer}{Peng
  et~al.}{2009}]{Peng2009}
Peng E.~W.,  Jord\'{a}n A.,  Blakeslee J.~P.,  Mieske S.,  C\^{o}t\'{e} P.,
  Ferrarese L.,  Harris W.~E.,  Madrid J.~P.,    Meurer G.~R.,  2009,
  \apj, 703, 42


\bibitem[Schlafly 
\& Finkbeiner(2011)]{Schlafly2011} Schlafly, E.~F., \& Finkbeiner, D.~P.\ 2011, \apj, 737, 103 


\bibitem[\protect\citeauthoryear{Sharma, Bland-Hawthorn, Johnston \&
  Binney}{Sharma et~al.}{2011}]{Sharma2011}
Sharma S.,  Bland-Hawthorn J.,  Johnston K.~V.,    Binney J.,  2011, \apj, 730, 3

\bibitem[\protect\citeauthoryear{Strader, Brodie, Spitler \& Beasley}{Strader
  et~al.}{2006}]{Strader2006}
Strader J.,  Brodie J.,  Spitler L.,    Beasley M.,  2006, \aj, 132,
  2333

\bibitem[\protect\citeauthoryear{Strader, Romanowsky, Brodie, Spitler, Beasley,
  Arnold, Tamura, Sharples \& Arimoto}{Strader et~al.}{2011}]{Strader2011}
Strader J.  et al.,  2011, \apjs, 197, 33

\bibitem[\protect\citeauthoryear{Tamura, Sharples, Arimoto, Onodera, Ohta \&
  Yamada}{Tamura et~al.}{2006a}]{Tamura2006a}
Tamura N.,  Sharples R.~M.,  Arimoto N.,  Onodera M.,  Ohta K.,    Yamada Y.,
  2006a, \mnras, 373, 588

\bibitem[\protect\citeauthoryear{Tamura, Sharples, Arimoto, Onodera, Ohta \&
  Yamada}{Tamura et~al.}{2006b}]{Tamura2006b}
Tamura N.,  Sharples R.~M.,  Arimoto N.,  Onodera M.,  Ohta K.,    Yamada Y.,
  2006b, \mnras, 373, 601

\bibitem[\protect\citeauthoryear{Zepf \& Ashman}{Zepf \&
  Ashman}{1993}]{Zepf1993}
Zepf S.,  Ashman K.,  1993, \mnras, 264

\bibitem[\protect\citeauthoryear{Zhu}{Zhu et al.}{2014}]{Zhu2014} Zhu, L., et al.\ 2014, \apj, 792, 59 




\end{thebibliography}


\end{document}